\documentclass[twocolumn]{autart} 
\usepackage{graphicx}      % include this line if your document contains figures 
\usepackage{natbib}        % required for bibliography
\usepackage{nicefrac}
\usepackage{tikz}
\usepackage{soul}
\usepackage{mdframed}
\usepackage{framed}
\usepackage{amsmath,amssymb}
\usepackage{graphicx}   
\usepackage{comment}
\usepackage{xcolor}
\usepackage{lipsum}
\usepackage{amssymb}
\usepackage[utf8]{inputenc}
\usepackage[utf8]{inputenc}
\usepackage[english]{babel}
\usepackage{enumitem} 
\newcommand{\Rr}{\mathbb{R}}

\newcommand{\MAP}[3]{{#1}:{#2}\to\mathbb{R}^{#3}}      %% Mappings

\theoremstyle{plain}
\newtheorem{dfn}{Definition}
\newtheorem{Thm}{Theorem}

\newtheorem{pro}{Proposition}

\newtheorem{Asu}{Assumption}

\newtheorem{Rem}{Remark}

\def\rea{\mathbb{R}}
\begin{document}
 \pdfoutput=1
\begin{frontmatter}	
\title{  Energy-Based Control Approaches for Weakly Coupled Electromechanical Systems \thanksref{footnoteinfo} }
\thanks[footnoteinfo]{This paper was not presented at any IFAC 
meeting. Corresponding author J. M. A. Scherpen.}
\author[First]{N. Javanmardi}\ead{n.javanmardi@rug.nl},    % Add the 
\author[Second]{P. Borja}\ead{pablo.borjarosales@plymouth.ac.uk},               % e-mail address 
\author[Third]{M. J. Yazdanpanah}\ead{yazdan@ut.ac.ir},  % (ead) as shown
\author[First]{J. M. A. Scherpen}\ead{ j.m.a.scherpen@rug.nl} 

\address[First]{Jan C. Willems Center for Systems and Control, ENTEG, Faculty of Science and Engineering, University of Groningen, Groningen, The Netherlands}
\address[Second]{School of Engineering, Computing and Mathematics, University of Plymouth, Plymouth, United Kingdom }
\address[Third]{Control and Intelligent Processing Center of Excellence, School of Electrical and Computer Engineering, University of Tehran, Tehran, Iran}
\begin{abstract}
  This paper addresses the regulation and trajectory-tracking problems for two classes of weakly coupled electromechanical systems. To this end, we formulate an energy-based model for these systems within the port-Hamiltonian framework. Then,
 we employ Lyapunov theory and the notion of contractive systems to develop control approaches in the port-Hamiltonian framework.
 Remarkably, these control methods eliminate the need for solving partial differential equations or implementing any change of coordinates, and are endowed with a physical interpretation. We also investigate the effect of coupled damping on the transient performance and convergence rate of the closed-loop system. Finally, the applicability of the proposed approaches is illustrated in two applications of electromechanical systems via simulations.      \end{abstract}
\begin{keyword}
Electromechanical systems, Port-Hamiltonian systems, Trajectory tracking, Energy shaping, Coupled damping, Contractive systems.
\end{keyword}
\end{frontmatter}
%===============================================================================
\section{Introduction}
\begin{comment}
\topskip=40pt
\parskip=10pt
\parindent=0pt
\baselineskip=15pt
\end{comment}
\parskip=1pt
Electromechanical (EM) systems are developed and utilized in a wide range of industrial applications, including electric drives, magnetic levitation systems, motors, micro/nanoelectromechanical systems (MEMS/NEMS) (see \cite{crowder2019electric,dean2009applications,zhang2014electrostatic}). They consist of mechanical and electrical subsystems coupled through nonlinear dynamics, where both subsystems exchange energy. Based on the coupling dynamics, EM systems can be classified as strongly coupled (such as motors or generators) and weakly coupled (such as MEMS or magnetic levitation systems). Notably, these highly inherently nonlinear coupling dynamics may lead to instability and other practical issues in the systems. For instance, pull-in instability, characterized by a saddle-node bifurcation, is a well-known critical issue in MEMS actuators resulting in operational limitations on stabilization---see  \cite{1372553} and \cite{zhang2014electrostatic}. To address these practical issues, different control methods have been proposed to solve stabilization and trajectory tracking problems and extend the allowable travel range in  EM applications \cite{chu1994analysis,maithripala2005control,vo2021novel,zhang2022observer,zhang2015continuous}.

Motivated by the multi-domain nature of EM systems, energy-based modeling and control approaches, particularly the port-Hamiltonian (pH) framework, have been investigated in the related literature \cite{fujimoto2003trajectory,maithripala2003port,nunna2015constructive,ortega2001putting,rodriguez2003stabilization,rodriguez2000novel,ryalat2020dynamic,ryalat2020improved}. In this regard, the pH approach is suitable to propose a unified framework to model a wide range of EM systems. Moreover, systematic pH-based tracking and stabilization control methods can be applied to reshape the open-loop dynamics into desired ones, e.g., the passivity-based control (PBC) method known as Interconnection and Damping Assignment (IDA)---introduced in \cite{ortega2002interconnection}---has been implemented to address regulation problems in a wide range of applications, including underactuated mechanical systems and EM systems \cite{ortega2002stabilization,ortega2001putting,rodriguez2003stabilization}.
%In \cite{ortega2002interconnection}, a passivity-based control (PBC) method known as Interconnection and Damping Assignment (IDA) is introduced. This method has been implemented to address regulation problems in a wide range of applications, including underactuated mechanical systems and EM systems \cite{rodriguez2003stabilization,ortega2001putting,1024334}.
However, the traditional IDA-PBC formulation exhibits two shortcomings that might hamper its implementation in some EM applications: (i) the technique requires solving a set of partial differential equations (PDEs) referred to as matching equations, which represent the main bottleneck of the approach, as they might be unsolvable, or finding the exact closed-form solutions might be infeasible; (ii) this strategy is not suitable to solve the trajectory-tracking problem. To overcome (i), some studies, e.g., \cite{borja2016constructive,nunna2015constructive}, propose strategies to remove the PDEs from the control design in particular classes of systems. In particular, the approach proposed in \cite{nunna2015constructive} relies on dynamic extensions, while the strategy formulated in \cite{borja2016constructive} requires the passive output to be integrable. Concerning (ii), in \cite{fujimoto2003trajectory} the authors propose a tracking control approach via generalized canonical transformations. However, finding an adequate transformation may require solving a set of PDEs, leading to the same issues as in IDA-PBC. The notion of contractive pH systems is adopted to develop a tracking version of the IDA-PBC method for a class of pH systems, termed as timed IDA-PBC tracking method \cite{yaghmaei2017trajectory}.
%The IDA-PBC technique requires solving a set of partial differential equations (PDEs) referred to as matching equations. Such PDEs represent the main bottleneck in IDA-PBC, as they might be unsolvable, or finding closed-form solutions might be infeasible. To overcome this problem, \cite{nunna2015constructive} introduces the concept of algebraic solutions to the matching equations. The authors of the mentioned reference also propose a dynamic IDA-PBC method to preserve the pH structure and ensure equilibrium asymptotic stability without solving PDEs. Similarly, in \cite{borja2016constructive}, the authors propose a PBC method based on feeding back the integral of the passive output, which does not require solving PDEs. \pbr{The notion of contractive pH systems is adopted to develop a tracking version of the IDA-PBC method for a class of pH systems, termed as timed IDA-PBC tracking method \cite{yaghmaei2017trajectory}.}
% Moreover, an energy shaping approach, based on the PBC method and power shaping output, is presented in \cite{borja2016constructive} for stabilizing pH systems, thereby circumventing the requirement of solving the PDEs.

The lack of effective coupling between electrical and mechanical subsystems in weakly coupled EM systems leads to some challenging outcomes.  In \cite{maithripala2003port}, a control approach based on Casimir functions is proposed for MEMS actuators. Nonetheless, due to the weak coupling between subsystems, the transient performance of the system is unaffected by the proposed controller.
In \cite{ortega2001putting} and \cite{rodriguez2000novel}, it is shown that the weak coupling between subsystems in a magnetic levitation system prevents the stabilization of the desired equilibrium by assigning a desired energy function while preserving the interconnection matrix of the open-loop system. In this regard, the IDA-PBC approach is an alternative to enhance the coupling between the subsystems by modifying the natural interconnection matrix. In \cite{rodriguez2003stabilization}, a general control method using IDA-PBC for addressing asymptotic regulation problems in a class of EM systems is proposed. However, the suggested approach involves finding a solution to the matching equations. To overcome this, in \cite{ryalat2020dynamic}, a dynamic control law based on the IDA-PBC approach employing a coordinate transformation is suggested. This approach improves the coupling between subsystems and solves the regulation problem for a class of weakly coupled EM systems. Concerning techniques to improve the closed-loop performance via damping injection, \cite{9804759} introduces the concept of coupled-damping injection and delves into the impact of gyroscopic forces and
coupled damping on the stability and performance of mechanical systems.

This paper addresses both regulation and trajectory tracking problems for two broad classes of weakly coupled EM systems within the pH framework. 
In particular, the control methods do not require solving PDEs or any change of coordinates. Moreover, these methods adopt static control design, i.e., they are formulated without dynamic extensions. Therefore,
the control structure is less complex compared to strategies relying on dynamic controllers and offers simplicity in terms of design and analysis.
Besides, we investigate coupled-damping injection and its impact on the stability and performance of the closed-loop system. The main contributions of the paper are summarized as follows: 

\begin{itemize}
\item Adopting a pH modeling approach, we propose a unified framework for controlling weakly coupled EM systems. The resulting controllers are static and do not require solving PDEs or changing the coordinates, easing their implementation in comparison with the methods reported in \cite{nunna2015constructive, rodriguez2003stabilization, ryalat2020dynamic}.
\item We provide constructive control design strategies that are suitable to solve the regulation and trajectory-tracking problems for some classes of weakly coupled EM systems. Note that the approaches adopted in \cite{nunna2015constructive, rodriguez2003stabilization, ryalat2020dynamic} focus on addressing the set-point regulation problem.
\item We show how energy-shaping strategies can be tailored and combined with the concept of contractive systems to control EM systems, achieving exponential stability results. Notably, employing a traditional
Lyapunov analysis instead of a contraction analysis leads to more complex proofs and conservative results.
\item We introduce the concept of coupled damping in EM systems and explore its potential to enhance the performance of the transient response and convergence rate.
\end{itemize}

%\pbr{@I would adapt this short paragraph to be a bullet contained on the list above}  We propose a pH framework for weakly coupled EM systems. The design procedure of the proposed control methods is based on the contractive pH systems and the power shaping output in the pH systems. Notably, the proposed trajectory-tracking control strategies guarantee exponential convergence to the desired trajectory.

Compared to \cite{rodriguez2003stabilization}, the proposed control methods remove the need to solve PDEs. Moreover, in contrast to the results reported in \cite{nunna2015constructive,ryalat2020dynamic}, where dynamic extensions are needed, the proposed controllers are static.
Besides, compared with all mentioned works, which only investigate the regulation problem, the current work elaborates on both regulation and trajectory tracking approaches. We emphasize that the timed IDA-PBC approach introduced in \cite{yaghmaei2017trajectory} presents a general tracking method using a set of PDEs applicable to a class of pH systems. However, it does not specifically address the stabilization problem, avoiding the need to solve PDEs and related control challenges for weakly coupled EM systems.                        
In the current work, we propose target dynamics and energy function in a way that avoids the need for solving PDEs or employing any change of coordinates for weakly coupled EM systems. The proposed approach also enhances the weak coupling between subsystems by proposing coupled damping terms, resulting in improving the performance of the closed-loop system.

 The remainder of the paper is organized as follows: Section \ref{sec:Review} provides a brief review of contractive pH systems. Section \ref{sec:EM moddel} introduces the class of weakly coupled EM pH systems under study, along with the problem formulation. In Section \ref{sec:reg}, we address the regulation problem for the class of EM systems introduced in Section \ref{sec:EM moddel}. We propose two trajectory-tracking approaches for these systems in Section \ref{sec:track}.
 In Section \ref{sec:coupleddamp}, we study the effect of coupled damping on the transient performance of the closed-loop system. The performance of the proposed methods in sections \ref{sec:reg} and \ref{sec:track} are simulated for MEMS and Maglev systems in Section \ref{sec:ex}. This study is summarized and discussed with conclusions in Section \ref{sec:con}. 

{\it Notation.} In the subsequent sections, $A\succ0$ ($A\succeq0$) means that the matrix A is positive definite (positive semi-definite), respectively. $\nabla H$ is defined as $[\,\frac{\partial H}{\partial x_1}, \frac{\partial H}{\partial x_2}, ... \frac{\partial H}{\partial x_n}]\,^\top$ for a continuously differentiable function $H:\mathbb{R}^n \longrightarrow \mathbb{R}$, and ${\nabla}^2 H$ is a matrix whose $ij$th element is $\frac{{\partial}^2 H}{\partial x_i \partial x_j}$. For a full-rank matrix $g \in \Rr^{n \times m}$, we define $g^{\dagger}\triangleq(g^{\top}g)^{-1}g^{\top}$. $I \in \Rr^{n \times n}$ represents the $n$-dimensional identity matrix.
%-------- Preliminaries -----------%
%%%%%%%%%%%%%%%%%%%%%%%%%%%%%%%%%%%
\section{Preliminaries}
\label{sec:Review}
\subsection{Contractive pH systems}
Consider the input-state-output representation of pH systems given by 
\begin{equation}
\label{eq:PHOL}
\begin{array}{rcll}
\dot{x}&=&(\mathcal{J}(x)-\mathcal{R}(x))\nabla \mathcal{H}(x) + g(x) u,\quad & x\in D_{\tt 0}\subseteq \Rr^n,\\
y&=&g^\top(x)\nabla \mathcal{H}(x),\quad &u,y \in \Rr^m,\vspace{-5 mm}
\end{array}
\vspace{3 mm}
\end{equation}

where $x(t)$ is the state, the interconnection matrix $\mathcal{J}: \Rr^n \to \Rr^{n\times n}$ is skew-symmetric, the damping matrix  $\mathcal{R}:  \Rr^n \to \Rr^{n\times n} $ is positive semi-definite, $\mathcal{H}: \Rr^n \to \Rr$ is the system's Hamiltonian, the input matrix $g: \Rr^n \to \Rr^{n \times m}$ satisfies $\mbox{rank}(g)=m \leq n$, $u,\ y$ are the input and output vectors, respectively, and $D_{\tt 0}$ is the state space of the system, which is an open subset of $\Rr^n$. For abbreviation, we define the matrix $\mathcal{F}:\Rr^{n} \to \Rr^{n \times n}$, $\mathcal{F}(x)\triangleq \mathcal{J}(x)-\mathcal{R}(x)$.

The following definition is necessary to present the technical content of this and the subsequent sections.

\begin{dfn}
\label{def:feasible}
$x^\star:\mathbb{T}\rightarrow\Rr^n$, where $\mathbb{T} \subset \Rr_{\tt +}$ , is said to be a \textit{feasible trajectory} of the system \eqref{eq:PHOL} if there exists $u^\star:\mathbb{T}\rightarrow \Rr^m$ such that, for all $t\in \mathbb{T}$, the following relation holds:
\begin{equation*}
\dot{x}^\star=\mathcal{F}(x^\star(t))\nabla \mathcal{H}(x^\star(t))+g(x^\star(t))u^\star(t).
\end{equation*}
 \end{dfn}

To tackle the trajectory tracking problem, we use the convergence property of {contractive} systems, which makes all the trajectories of the systems converge exponentially together as $t\to \infty$. Therefore, the closed-loop system is designed such that it is contractive, while the desired trajectory $x^{\star}$ is a {feasible trajectory} of the system. In particular, the following theorem provides the result on contractive systems that represents the foundation of the tracking results reported in this paper. The mentioned theorem is taken from \cite{yaghmaei2017trajectory}, where the authors also introduce the timed IDA-PBC method, which is a contraction-based, trajectory-tracking approach for a class of pH systems. See \cite{yaghmaei2017trajectory} for further details on the timed IDA-PBC method and the proof of the next theorem.
\begin{Thm}[\cite{yaghmaei2017trajectory}]
	\label{th:contractive}
	Consider the following system
	\begin{equation}
	\label{closed sys}
% 	\begin{split}
	\dot{x}=\mathcal{F}_{\tt d} \nabla \mathcal{H}_{\tt d}(x, t),
% 	\end{split}
	\end{equation}
	with $\mathcal{F}_{\tt d}\triangleq \mathcal{J}_{\tt d}-\mathcal{R}_{\tt d}$, where $\mathcal{J}_{\tt d}=-\mathcal{J}^\top_{\tt d}$ and $\mathcal{R}_{\tt d}=\mathcal{R}^\top_{\tt d}\succeq 0$ are the desired constant interconnection and damping matrices, respectively. The system \eqref{closed sys} is contractive on the open subset $D_0 \subseteq \Rr^n $ if:
	\begin{itemize}
	 \item [(i)] All the eigenvalues of $\mathcal{F}_{\tt d}$ have strictly negative real part.
	 \item [(ii)] The desired Hamiltonian function $\mathcal{H}_{\tt d}: \Rr^n \times \Rr_+  \to \Rr$ satisfies
	\begin{equation}
	\label{cond3}
	\begin{split}
	&\gamma_1 I\prec \nabla^2 \mathcal{H}_{\tt d}(x, t) \prec \gamma_2 I,\quad \forall x \in D_0,
 \end{split}
\end{equation}
   for constants $\gamma_1,\gamma_2$, such that $0<\gamma_1<\gamma_2$.
	 \item [(iii)] There exists a positive constant $\varepsilon$ such that
	\begin{equation}
	\label{Eq:N}
	Q \triangleq \begin{bmatrix}
	\mathcal{F}_{\tt d} & \left( 1-\frac{\gamma_1}{\gamma_2} \right) \mathcal{F}_{\tt d}\mathcal{F}_{\tt d}^\top\\ -\left(1-\frac{\gamma_1}{\gamma_2}+\varepsilon\right)I & -\mathcal{F}_{\tt d}^\top
	\end{bmatrix},
	\end{equation}
	has no eigenvalues on the imaginary axis.
	\end{itemize}
\end{Thm}

\section{Modeling and Problem Formulation}
\label{sec:EM moddel}
EM systems consist of the interconnection of mechanical and electrical subsystems. We consider that the states of the mechanical part are given by the generalized position $q\in\rea^{n_{\tt m}}$ and momenta $p\in\rea^{n_{\tt m}}$, while the state of the electrical subsystem is given by $x_{\tt e}\in\rea^{n_{\tt e}}$. To simplify the notation, we denote the states of the whole system as
\begin{equation}
\eta^\top = \begin{bmatrix}
    q^\top & \; &  p^\top  & \; & x^\top_{\tt e}
\end{bmatrix}.
\end{equation}

We restrict our attention to weakly coupled EM systems that admit a pH representation of the form
\begin{equation}
\label{eq:open-loop elecmech}
\arraycolsep=1pt
\def\arraystretch{1.2}
\begin{array}{rcl}
 \begin{bmatrix}
  \dot{q} \\ \dot{p} \\ \dot{x}_{\tt e}
 \end{bmatrix}&=& \begin{bmatrix}0 & \quad  I & \quad  0 \\ -I & \quad -R_{\tt m} & \quad 0 \\0 & \quad  0 & \quad  J_{\tt e}-R_{\tt e}\end{bmatrix}\begin{bmatrix}
{{\nabla _{{q}}}{\mathcal{H}(\eta)}} \\
{{\nabla _{{p}}}{\mathcal{H}(\eta)}} \\
{{\nabla _{{x}_{\tt e}}}{\mathcal{H}(\eta)}}
\end{bmatrix} + \begin{bmatrix}
 0 \\0\\ G_{\tt e}
\end{bmatrix}u,\\ [0.7cm]\mathcal{H}(\eta)&=& \displaystyle\frac{1}{2}{p^{\top}} M^{-1}(q) p+V(q)+H_{\tt e}(q,x_{\tt e}), \vspace{-5 mm}
\end{array}
\end{equation}
\vspace{.5 mm}

where $\MAP{M}{\rea^{n_{\tt m}}}{{{n_{\tt m}}\times {n_{\tt m}}}}$ is the mass inertia matrix, which is positive definite; $\MAP{V}{\rea^{n_{\tt m}}}{}$ is the mechanical potential energy; $\MAP{H_{\tt e}}{\rea^{n_{\tt m}}\times \rea^{n_{\tt e}}}{}$ is the coupling energy; $R_{\tt m}\in\rea^{n_{\tt m} \times n_{\tt m}}$ is the damping matrix of the mechanical subsystem, which is positive semi-definite; $J_{\tt e}\in\rea^{n_{\tt e} \times n_{\tt e}}$ is the interconnection matrix of the electrical subsystem, which is skew-symmetric; $R_{\tt e}\in\rea^{n_{\tt e} \times n_{\tt e}}$ is the damping matrix of the electrical subsystem, which is positive semi-definite; $G_{\tt e}\in\rea^{n_{\tt e} \times n_{\tt e}}$ is the full-rank input matrix; $u\in\Rr^{n_{\tt e}}$ denotes the input vector. In particular, we consider that the electrical energy-storing elements are capacitors or inductors whose capacitance or inductance depends on the mechanical position $q$. Hence, we consider the following structure for the coupling energy:
\begin{equation}
\label{eq:elec energy}
 H_{\tt e}(q,x_{\tt e})=\frac{1}{2}x_{\tt e}^\top 	\Psi (q)x_{\tt e},
\end{equation}
where $\MAP{\Psi}{\rea^{n_{\tt m}}}{n_{\tt e} \times n_{\tt e}}$ is the capacitance or inductance matrix, which is positive definite. We stress that physical limitations often constrain the range of motion---i.e., the possible values for $q$---in EM systems.

\begin{Rem}
 Customarily, the electrical and mechanical subsystems of an EM device are coupled through their energy via nonlinear functions. In particular, these nonlinear couplings may lead to the so-called pull-in instability, a saddle-node bifurcation phenomenon, which is one of the critical practical problems in MEMS devices. As a result, performance and operation range in MEMS are significantly limited by the inherent instability of these systems \cite{maithripala2003capacitive,zhang2014electrostatic}.
\end{Rem}

The model \eqref{eq:open-loop elecmech} is suitable for representing a broad range of weakly coupled EM devices, including electrostatic MEMS/NEMS and magnetic levitation systems (see, e.g., \cite{maithripala2003capacitive,rodriguez2003stabilization,rodriguez2000novel}).  Therefore, developing constructive control approaches for \eqref{eq:open-loop elecmech} is highly relevant. This paper focuses on two different control problems, which are described below.

\begin{itemize}
    \item {\bf Set-point regulation.} The set of assignable equilibria for \eqref{eq:open-loop elecmech} is given by
    \begin{equation*}
        \mathcal{E} = \left\lbrace (q,x_{\tt e})\in\rea^{n_{\tt m}}\times\rea^{n_{\tt e}}\mid \nabla_{q}\mathcal{H}(\eta) = \mathbf{0} ; p=\mathbf{0} \right\rbrace.
    \end{equation*}
 Accordingly, for every configuration $\bar\eta\in\mathcal{E}$, there exists $\bar{u}\in\rea^{n_{\tt e}}$ such that $\bar{\eta}$ is an equilibrium for \eqref{eq:open-loop elecmech} in closed-loop with $u=\bar{u}$. Hence, the regulation (stabilization) problem consists in finding a control law $u$ such that the trajectories of the closed-loop system converge to the desired equilibrium
 \begin{equation}
 \label{eq:desired eq}
     \eta^\top_{\tt d}\triangleq\begin{bmatrix}
         q^\top_{\tt d} & \; &  p^\top_{\tt d} & \; &  x^\top_{\tt e_{d}}
     \end{bmatrix},
 \end{equation}
where $\eta_{\tt d}\in\mathcal{E}$, and $q_{\tt d}, \,  p_{\tt d}\,\,\, \mbox{and,} \,\, x_{\tt e_{d}}$ denote the desired position, momenta, and electrical states, respectively.
 \item {\bf Trajectory tracking.} Consider the desired trajectory
 \begin{equation}
 \label{eq:desired trajectory}
 {\eta^\star}^\top(t)\triangleq\begin{bmatrix}
         {q^{\star}}^\top(t) & \; &  {p^\star}^\top (t) & \; &  {x_{\tt e}^\star}^\top(t)
     \end{bmatrix},
 \end{equation}
 which is a feasible trajectory---see Definition \ref{def:feasible}---where $ q^{\star}(t),\,\,\,p^\star (t)\,\,\, \mbox{and}\,\,\, x_{\tt e}^\star(t)$ in \eqref{eq:desired trajectory} represent the desired position, momenta, and electrical states trajectories, respectively.
 The trajectory-tracking problem consists in finding a control law $u$ such that the trajectories of the closed-loop system track $\eta^\star(t)$. To simplify the notation, we omit the argument $t$ from the desired trajectory $\eta^\star$ in the subsequent sections.
\end{itemize}

To ease the presentation of the results, we define the following mappings:
\begin{eqnarray}
\bar{R}_{\tt e}&\triangleq& R_{\tt e}+K_{\tt e}\succ 0, \label{barRe} \\
 z &\triangleq& -G_{\tt e}^{\top}\left(J_{\tt e}-\bar{R}_{\tt e}\right)^{-\top}\left(\Gamma^{\top}q + x_{\tt e}  \right), \label{z} \\
 z_{\tt d} &\triangleq& -G_{\tt e}^{\top}\left(J_{\tt e}-\bar{R}_{\tt e}\right)^{-\top}\left(\Gamma^{\top}q_{\tt d} + x_{\tt e_{d}}  \right), \label{zd} \\
 z^{\star} &\triangleq& -G_{\tt e}^{\top}\left(J_{\tt e}-\bar{R}_{\tt e}\right)^{-\top}\left(\Gamma^{\top}q^{\star} + x^{\star}_{\tt e}  \right), \label{zstar}
\\
 \Upsilon_1 &\triangleq& \left.\nabla_q \left(\nabla_{x_{\tt e}}H_{\tt e}(q,x_{\tt e})\right)\right|_{(q,x_{\tt e})=(q_{\tt d},x_{\tt e_{d}})}, \label{Upsilon}
\end{eqnarray}
where $\Gamma\in\rea^{n_{\tt m}\times n_{\tt e}}$ is a constant matrix and $K_{\tt e}\in\rea^{n_{\tt e} \times n_{\tt e}}$ is a positive semi-definite matrix such that \eqref{barRe} holds and the matrix $J_{\tt e}-\bar{R}_{\tt e}$ is invertible.
\subsection{Closed-loop system with coupled damping}
\label{sec:coupledamping}
%Here, we introduce the closed-loop system utilized in the next sections as follows:
We consider a target closed-loop system with the following structure
\begin{equation}
\label{eq:closedloop}
 \arraycolsep=1pt
\def\arraystretch{1.2}
\begin{array}{rcl}
 \begin{bmatrix}
  \dot{q} \\ \dot{p} \\ \dot{x}_{\tt e}
 \end{bmatrix}&=& \begin{bmatrix}
                 0 & \quad  I & \quad  0 \\ -I & \quad -R_{\tt m} & \quad \Gamma \\
                 0 & \quad  -\Gamma^\top+D_{\tt d} & \quad  J_{\tt e}-{\bar{R}_{\tt e}}                 \end{bmatrix}\begin{bmatrix}
                 {{\nabla _{{q}}}{\mathcal{H}_{\tt d}}} \\
	             {{\nabla _{{p}}}{\mathcal{H}_{\tt d}}} \\
	             {{\nabla _{{x}_{\tt e}}}{\mathcal{H}_{\tt d}}}
                 \end{bmatrix},
\end{array} 
\end{equation}

where $\bar{R}_{\tt e}\in \rea^{n_{\tt e} \times n_{\tt e}}$ represents the desired electrical damping, which is positive definite. Besides, $D_{\tt d} \in \rea^{n_{\tt e} \times n_{\tt m}}$ denotes the coupled damping term, which couples the mechanical and electrical subsystems. Accordingly, the coupling between electrical and mechanical subsystems can be strengthened by injecting coupled damping. The desired Hamiltonian function $\mathcal{H}_{\tt d}$ is precisely defined for the regulation and tracking cases in Sections \ref{sec:reg} and \ref{sec:track}, respectively.

A more compact notation for \eqref{eq:closedloop} is given by
%We denote \eqref{eq:closedloop} briefly by the following formulation:
	\begin{equation}
	\label{eq:closedJ-R}
	\dot{\eta}=\mathcal{F}_{\tt d} \nabla \mathcal{H}_{\tt d},
	\end{equation}
where $\mathcal{F}_{\tt d}=\mathcal{J}_{\tt d}-\mathcal{R}_{\tt d}$  with the following matrices:

\begin{equation}
\label{Eq:R-J}
\begin{array}{rcl}
	\mathcal{J}_{\tt d}& \triangleq &
 \begin{bmatrix}
	\mathbf{0} & I & \mathbf{0}\\[1 mm] -I & \mathbf{0} & \Gamma-\frac{1}{2}D^\top_{\tt d}\\[1 mm]\mathbf{0} & -\Gamma^\top+\frac{1}{2}D_{\tt d} & J_{\tt e}\end{bmatrix}, \\[0.6cm] \mathcal{R}_{\tt d} &\triangleq &\begin{bmatrix}
	\mathbf{0} & & \mathbf{0} & &\mathbf{0}\\[1 mm] \mathbf{0} & &R_{\tt m} & &-\frac{1}{2} D^\top_{\tt d}\\[1 mm]\mathbf{0} & &-\frac{1}{2} D_{\tt d}&&\bar{R}_{\tt e}
\end{bmatrix}.
\end{array}
\end{equation}

Hence, \eqref{eq:closedloop} and \eqref{eq:closedJ-R} correspond to a pH system,
where $\mathcal{J}_{\tt d}$ is skew-symmetric and $\mathcal{R}_{\tt d}$ is symmetric and positive semi-definite, if the following condition holds for the matrices $D_{\tt d}$ and $\bar{R}_{\tt e}$:
\begin{equation}
\label{eq:pHcondition}
     {R}_{\tt m}-\frac{1}{4}D^\top_{\tt d}\bar{R}^{-1}_{\tt e}D_{\tt d}\succeq 0.
\end{equation}

%%%%%%%%%%%%%%%%%%%%%%%%%%%%%%%%%%%%%
%----------- Main Results -----------%
%%%%%%%%%%%%%%%%%%%%%%%%%%%%%%%%%%%
\section{Set-point Regulation Control Design}
\label{sec:reg}
This section provides constructive energy-shaping control approaches to stabilize two specific classes of weakly coupled EM systems. The proposed control methods do not require solving partial differential matching equations or implementing any change of coordinates. Moreover, these methods employ static control design.
\subsection{Regulation for EM systems with only partial mechanical damping}
\label{sec:reg1}
The next assumption introduces the class of EM systems studied in this section.
\begin{Asu}
\label{ass:class1}
  The system \eqref{eq:open-loop elecmech} satisfies $\nabla^{2}_{q}\mathcal{H}(\eta_{\tt d})\succ 0$ and  $R_{\tt m}\succ0 $.
\end{Asu}

From a physical perspective, Assumption \ref{ass:class1} implies that the mechanical subsystem is fully damped. Moreover, the condition imposed on the energy function guarantees that, after assigning the desired equilibrium, the resulting closed-loop energy function is convex with respect to this point without changing the natural interconnection matrix of the system. Remarkably, a broad range of EM systems, including MEMS, microphones, or loudspeaker systems \cite{javanmardi2023contraction}, satisfies this assumption. In \cite{borovic2004control}, the authors model and propose a control strategy to stabilize a physical device satisfying Assumption \ref{ass:class1}. For systems with a linear function $\Psi$ in \eqref{eq:elec energy} and a constant mass inertia matrix, the assumption simplifies to a stable and strictly convex---with respect to the desired configuration---potential energy function, denoted by $\nabla^{2}_{q}V(q_{\tt d})\succ 0$.

%This assumption is applicable to EM systems characterized by a stable convex energy function, expressed as $\nabla^{2}_{q}\mathcal{H}(\eta_{\tt d})\succ 0$. For systems with a linear function $\Psi$ in \eqref{eq:elec energy} and a constant mass inertia matrix, the assumption simplifies to a stable and convex potential energy function, denoted by $\nabla^{2}_{q}V(q_{\tt d})\succ 0$. Additionally, the natural mechanical damping $R_{\tt m}$ is positive definite in this class of EM systems. It is worth noting that a broad range of weakly coupled EM systems, including MEMS, microphones, or loudspeaker systems \cite{javanmardi2023contraction}, satisfies this assumption.

The following proposition provides a constructive control strategy to solve the regulation problem for EM systems satisfying Assumption \ref{ass:class1}.
%Now, we develop the control approach to solve the regulation problem for the EM systems in Assumption \ref{ass:class1}.
\begin{pro}\label{pro:case1}
Suppose that \eqref{eq:open-loop elecmech} satisfies Assumption \ref{ass:class1}. Consider a twice differentiable function $\MAP{\Phi_1}{\rea^{n_{\tt e}}}{}$, $K_{\tt e} \succeq 0$, and
the coupled damping matrix $D_{\tt d}$ such that \eqref{eq:pHcondition} holds with a strict inequality, and
\begin{eqnarray}
\nabla_{x_{\tt e}}H_{\tt e}(q_{\tt d},x_{\tt e_{d}})+\nabla \Phi_1(x_{\tt e_{d}}) = \mathbf{0},\label{phi11}\\
\begin{bmatrix}
\nabla_q^{2}\mathcal{H}(\eta_{\tt d})   %+\nabla^2_{x_{\tt e}}H_{\tt e}(q_{\tt d},x_{\tt e_{d}})%
& \; & \Upsilon_1 \\ \Upsilon_1^\top & \; & \nabla^{2}_{x_{\tt e}}H_{\tt e}(q_{\tt d},x_{\tt e_{d}})+\nabla^{2} \Phi_1(x_{\tt e_{d}})\end{bmatrix}\succ 0,
 \label{phi12}
\end{eqnarray}
where $\Upsilon_1 $ is defined in \eqref{Upsilon}. The control law
\begin{equation}
\begin{array}{rcl}
\label{eq:c1}
 u &=& G_{\tt e}^{-1}\left\lbrace \left( J_{\tt e}-\bar{R}_{\tt e} \right)\nabla \Phi_{1}(x_{\tt e}) - K_{\tt e}\Psi(q)x_{\tt{e}}
  \right.
\\&&+\left.D_{\tt d} M^{-1}(q)p\right\rbrace,
 \end{array}
\end{equation}
renders the desired equilibrium (locally) asymptotically stable.
\end{pro}

\begin{pf*}{Proof}
Given the controller \eqref{eq:c1}, the closed-loop system takes the form in \eqref{eq:closedloop} with
\begin{equation*}
\label{eq:energyrug}
\Gamma=\mathbf{0}, \quad \mathcal{H}_{\tt d}(\eta) = \mathcal{H}(\eta) + \Phi_{1}(x_{\tt e})+c_1,
\end{equation*}
where $c_1\triangleq -\mathcal{H}(\eta_{\tt d}) - \Phi_1(x_{\tt e_{d}}).$
%\begin{equation}
 %c= -H(q_{\star},\mathbf{0},x_{\tt e_{\star}}) - \Phi_1(x_{\tt e_{\star}}).
%\end{equation}
Hence,
\begin{equation}
 \mathcal{H}_{\tt d}(\eta_{\tt d}) = 0. \label{eq:Hd}
\end{equation}

Furthermore, some simple computations---omitted due to space constraints---show that \eqref{phi11} and \eqref{phi12} imply
\begin{equation}
 \nabla \mathcal{H}_{\tt d}(\eta_{\tt d}) = \mathbf{0}, \quad \nabla^{2} \mathcal{H}_{\tt d}(\eta_{\tt d}) \succ 0, \label{eq:gradHd}
\end{equation}
in a neighborhood of the desired equilibrium. Note that \eqref{eq:Hd} and \eqref{eq:gradHd} imply that $\mathcal{H}_{\tt d}(\eta)$ is positive with respect to $\eta_{\tt d}$ in the mentioned neighborhood. Additionally,
\begin{equation}
    \dot{\mathcal{H}}_{\tt d} = -\mathcal{Z}^\top \mathcal{B}_{\tt T}\mathcal{Z},
 \label{dHd}
\end{equation}

where
\begin{eqnarray}
\label{eq:Beta}
\mathcal{Z}&\triangleq& \begin{bmatrix} \nabla_{p}\mathcal{H}^\top_{\tt d} & \; & \nabla_{x_{\tt e}}\mathcal{H}^\top_{\tt d}\end{bmatrix}^\top,\\[1 mm]
\mathcal{B}_{\tt T}&\triangleq&\begin{bmatrix}
R_{\tt m}&-\frac{1}{2}D^\top_{\tt d}\\[1 mm]-\frac{1}{2}D_{\tt d}&\bar{R}_{\tt e}
\end{bmatrix}\label{Bt}.
\end{eqnarray}

According to the Schur complement of $\mathcal{B}_{\tt T}$, $R_{\tt m}  \succ 0$, $\bar R_{\tt e}  \succ 0$, and \eqref{eq:pHcondition}---with a strict inequality---ensure that $\mathcal{B}_{\tt T} \succ 0$.
Hence, $\dot{\mathcal{H}}_{\tt d} \leq 0$ and $\mathcal{H}_{\tt d}(\eta)$ is positive with respect to the desired equilibrium and non-increasing. Consequently, from Lyapunov's theory, $\eta_{\tt d}$ is a (locally) stable equilibrium point for the closed-loop system with Lyapunov function $\mathcal{H}_{\tt d}(\eta)$.

To prove asymptotic stability, note that \eqref{eq:gradHd} implies that $\eta_{\tt d}$ is a strict minimum of $\mathcal{H}_{\tt d}(\eta)$. Consequently, there exists a neighborhood of the equilibrium, denoted as $\mathcal{U}$, such that
\begin{equation}
    \nabla \mathcal{H}_{\tt d}(\eta) = \mathbf{0} \iff \eta = \eta_{\tt d} \quad \forall \eta\in\mathcal{U}. \label{asym11}
\end{equation}

%From ${R}_{\tt m}\succ 0$ and $\bar{R}_{\tt e}\succ 0$, we obtain
From \eqref{dHd}, $\dot{\mathcal{H}}_{\tt d}\equiv 0$ has the following implications
\begin{equation*}  \mathcal{Z}^\top \mathcal{B}_{\tt T}\mathcal{Z} = 0 \implies \mathcal{Z} = 0 \implies  \left\lbrace \begin{array}{l}
\nabla_{p}\mathcal{H}_{\tt d} = p = \mathbf{0}, \\
\nabla_{x_{\tt e}}\mathcal{H}_{\tt d} = \mathbf{0}.
   \end{array} \right. \label{dHd11}
\end{equation*}
%\begin{equation*}
%   \dot{\mathcal{H}}_{\tt d}\equiv 0 \implies \left\lbrace \begin{array}{rcl}
 %      R_{\tt m}\nabla_{p}\mathcal{H}_{\tt d}= \mathbf{0} & \implies & \nabla_{p}\mathcal{H}_{\tt d} = p = \mathbf{0} \\
  %     \bar{R}_{\tt e}\nabla_{x_{\tt e}}\mathcal{H}_{\tt d}= \mathbf{0} & \implies & \nabla_{x_{\tt e}}\mathcal{H}_{\tt d} = \mathbf{0}
   %\end{array} \right. \label{dHd11}
%\end{equation*}

Moreover,
\begin{equation}
    p= \mathbf{0} \implies \dot{p}= \mathbf{0} \implies \nabla_{q}\mathcal{H}_{\tt d}= \mathbf{0}. \label{dHd12}
\end{equation}

Hence, by combining \eqref{asym11} and \eqref{dHd12}, we get that $\dot{\mathcal{H}}_{\tt d}\equiv0$ implies $\nabla \mathcal{H}_{\tt d}(\eta_{\tt d}) = \mathbf{0}$. Thus, since the desired equilibrium is stable, it follows from LaSalle's invariance principle that the trajectories starting on $\mathcal{U}$ converge to $\eta_{\tt d}$. $\square$ 
\end{pf*}
%Remarkably, the transient performance of EM systems is governed by the mechanical subsystem. Accordingly, it is crucial to provide an adequate mechanical damping injection to shape the transient performance and optimize the convergence rate in the closed-loop system. The control approach discussed in Proposition \ref{pro:case1} employs the coupled damping $D_{\tt d}$ to ameliorate the coupling between mechanical and electrical subsystems. In particular, this coupling is achieved by modifying the desired interconnection matrix. Therefore, damping is appropriately injected into the mechanical subsystem from the electrical subsystem. Notably, the method proves advantageous in addressing the regulation problem for the weakly coupled EM systems characterized by non-zero and low natural mechanical damping $R_{\tt m}$.
\subsection{Regulation for EM systems with only partial or no mechanical damping}

Some weakly coupled EM systems, e.g., magnetic levitation systems, do not satisfy Assumption \ref{ass:class1} because the mechanical subsystem is only partially damped or undamped. Similarly, the potential energy of the system might not satisfy the conditions imposed in the mentioned assumption. To address this issue, the following assumption characterizes a class of EM systems whose coupling and mechanical properties---damping and potential energy---differ from those studied in Subsection \ref{sec:reg1}.

%In the next part, we turn to another class of weakly coupled EM systems introduced in the following assumption.
\begin{Asu}
\label{ass:class2}
Given \eqref{eq:open-loop elecmech}, there exists a twice differentiable function $\MAP{\varphi_1}{\rea^{n_{\tt e}}}{}$ such that
\begin{equation}
\begin{array}{rcl}
 \nabla_{q}H_{\tt e}(q,x_{\tt e}) &=& -\Gamma\nabla\varphi_1(x_{\tt e}), \\
 \nabla^{2}\varphi_1(x_{\tt e_{d}})&\succ& 0.
\end{array} \label{assHereg}
\end{equation}
\end{Asu}

Note that Assumption \ref{ass:class2} does not impose conditions on the mechanical damping. Therefore, we can relax the condition $R_{\tt m}\succ 0$ by considering $R_{\tt m}\succeq 0$. Furthermore, \eqref{assHereg} implies that
 the coupling energy is linear with respect to $q$. As mentioned, a relevant class of EM systems satisfying the assumption above includes magnetic levitation systems. The following proposition provides a controller that stabilizes EM systems characterized by Assumption \ref{ass:class2}.

\begin{pro}\label{pro:case2}
Consider the system \eqref{eq:open-loop elecmech} satisfying Assumption \ref{ass:class2} and matrices $K_{\tt e}\succeq 0$, $D_{\tt d}$ satisfying  \eqref{eq:pHcondition}. Let $\MAP{\Phi_2}{\rea^{n_{\tt e}}}{}$ be a twice differentiable function  such that
\begin{eqnarray}
\setlength{\arraycolsep}{1pt}
\arraycolsep=1pt
\def\arraystretch{0.5}
    &&\begin{bmatrix}
        \nabla_q \Phi_2(z_{\tt d}) \\
        \nabla_{x_{\tt e}} \Phi_2(z_{\tt d}) + \nabla\varphi_1(x_{\tt e_{\tt d}})
    \end{bmatrix} = \begin{bmatrix}
        \mathbf{0} \\ \mathbf{0} 
    \end{bmatrix},\label{Phic21} \\
   && \begin{bmatrix}
     \label{Phic22}
        \nabla_q^{2} \mathcal{H}(q_{\tt d})+
        \nabla_q^{2} \Phi_2(z_{\tt d})
        \hspace{-0.9cm}& \Upsilon_2 \\ \Upsilon^\top_2 & \nabla_{x_{\tt e}}^2 \Phi_2(z_{\tt d}) + \nabla^2\varphi_1(x_{\tt e_{d}})\end{bmatrix}\succ  0,\qquad
\end{eqnarray} 
where $\Upsilon_2\triangleq \left.\nabla_q\left(\nabla_{x_{\tt e}}\Phi_{2}(z)\right)\right|_{z = z_{\tt d}}$
, and $z$ and $z_{\tt d}$ are defined in \eqref{z} and \eqref{zd}, respectively. Therefore,
the system \eqref{eq:open-loop elecmech} in closed-loop with the controller
\begin{equation}
\label{eq:u2}
 \begin{array}{rcl}
  u &=& G_{\tt e}^{-1}\left\lbrace \left( R_{\tt e}-J_{\tt e} \right)\Psi(q)x_{\tt{e}}+(D_{\tt d}- \Gamma^\top) M^{-1}(q)p\right. \\ &&+\left.\left(  J_{\tt e}-\bar{R}_{\tt e}\right)  \nabla\varphi_1(x_{\tt e})\right\rbrace-\nabla_{z}\Phi_2(z), \vspace{-5 mm}
 \end{array}
 \vspace{3 mm}
\end{equation}

has a (locally) stable equilibrium at $\eta_{\tt d}$. Furthermore, the equilibrium is asymptotically stable if
\begin{equation}
    \nabla\varphi_1(x_{\tt e}) + \nabla_{x_{\tt e}}\Phi_{2}(z) = \mathbf{0} \implies \eta=\eta_{\tt d}. \label{detect}
\end{equation}
\end{pro}
%%%%%%%%%%%%%%%%%%

\begin{pf*}{Proof.}
From \eqref{z} and the chain rule, we have that
\begin{equation*}
   \begin{array}{rcl}
      \nabla_{q}\Phi_2  & = & - \Gamma(J_{\tt e}-\bar{R}_{\tt e})^{-1}G_{\tt e}\nabla_z\Phi_2,  \\
      \nabla_{x_{\tt e}}\Phi_2  & = & -(J_{\tt e}-\bar{R}_{\tt e})^{-1}G_{\tt e}\nabla_z\Phi_2.
   \end{array}
\end{equation*}

Hence,
\begin{equation}
    \begin{array}{rcl}
        \nabla_z\Phi_2 & = & -G_{\tt e}^{-1}(J_{\tt e}-\bar{R}_{\tt e})\nabla_{x_{\tt e}}\Phi_2,\\
         \nabla_{q}\Phi_2  &=& \Gamma\nabla_{x_{\tt e}}\Phi_2.
    \end{array}\label{nPhi2}
\end{equation}

Accordingly, applying the controller \eqref{eq:u2} results in the closed-loop system given by \eqref{eq:closedloop}, with
\begin{equation}
\label{eq:energyrug2}
\mathcal{H}_{\tt d}(\eta)\triangleq \frac{1}{2}p^{\top}M^{-1}(q)p + V(q) + \Phi_2(z) + \varphi_1(x_{\tt e}) + c_2,
\end{equation}
where $c_2 \triangleq - V(q_{\tt d}) - \Phi_2(z_{\tt d}) - \varphi_1(x_{\tt e_{ d}})$. Thus, from \eqref{nPhi2} and Assumption \ref{ass:class2}, we conclude that
\begin{equation}
\label{eq:matching}
 -\nabla_{q}\mathcal{H}_{\tt d} + \Gamma \nabla_{x_{\tt e}} \mathcal{H}_{\tt d} = -\nabla_{q}\mathcal{H}.
\end{equation}
where $\mathcal{H}_{\tt d}$ is defined in \eqref{eq:energyrug2}.
Consequently, the dynamics of $p$ are the same as in open loop. Note that this is equivalent to satisfying the matching equation in IDA-PBC.

The constant $c_2$ guarantees that \eqref{eq:energyrug2} evaluated at the desired equilibrium---i.e., $\mathcal{H}_{\tt d}(\eta_{\tt d})$--- is zero. Moreover, \eqref{Phic21} and \eqref{Phic22} imply that
\begin{equation*}
\nabla_{\eta}\mathcal{H}_{\tt d}(\eta_{\tt d})=\mathbf{0}, \qquad \nabla^{2}_{\eta}\mathcal{H}_{\tt d}(\eta_{\tt d})\succ 0.
\end{equation*}

Consequently, $\mathcal{H}_{\tt d}(\eta)$ is positive with respect to $\eta_{\tt d}$ in a neighborhood of the desired equilibrium. Additionally, the derivative of $\mathcal{H}_{\tt d}(\eta)$, along the trajectories of the system, takes the form \eqref{dHd}. 
From the Schur complement of $\mathcal{B}_{\tt T}$ in \eqref{Bt}, $R_{\tt m}  \succ 0$, $\bar R_{\tt e}  \succ 0$, and \eqref{eq:pHcondition} ensure that $\mathcal{B}_{\tt T} \succeq 0$.
Therefore,  $\dot{\mathcal{H}}_{\tt d} \leq 0$. Consequently, $\mathcal{H}_{\tt d}(\eta)$ is non-increasing. Thus, $\eta_{\tt d}$ is a (locally) stable equilibrium for the closed-loop system with Lyapunov function $\mathcal{H}_{\tt d}(\eta)$.
%$\mathcal{H}_{\tt d}(\eta)$ has a strict minimum at $\eta_{\tt d}$. Consequently, $\mathcal{H}_{\tt d}(\eta)$ is positive with respect to the desired equilibrium. By a similar argument in \eqref{dHd}-\eqref{eq:Beta} for $\mathcal{H}_{\tt d}(\eta)$ in \eqref{eq:energyrug2}, we conclude that $\dot{\mathcal{H}}_{\tt d}  \leq 0$  if and only if $\mathcal{B}_{\tt T} \succeq 0$. From Schur complement, $\mathcal{B}_{\tt T} \succeq 0$ if and only if \eqref{phic23} holds. Therefore, $\mathcal{H}_{\tt d}$  is non-increasing.
 %Thus, $\eta_{\tt d}$ is a stable equilibrium for the closed-loop system, and $\mathcal{H}_{\tt d}(\eta)$ qualifies as a Lyapunov function. 

Note that the mechanical system can be undamped. Therefore, we consider the worst-damping scenario,
i.e., $R_{\tt m}=\mathbf{0}$ and $\bar{R}_{\tt e}\succ 0$. Hence, 
from \eqref{dHd},
\begin{equation*}
 \dot{\mathcal{H}}_{\tt d}\equiv 0  \implies \nabla_{x_{\tt e}}\mathcal{H}_{\tt d} = \mathbf{0}
 \iff \nabla\varphi_1(x_{\tt e}) + \nabla_{x_{\tt e}}\Phi_{2}(z) = \mathbf{0}.
\end{equation*}

%\begin{equation*}
 %  \dot{\mathcal{H}}_{\tt d}\equiv 0 \implies \left\lbrace \begin{array}{rcl}
 %      R_{\tt m}\nabla_{p}\mathcal{H}_{\tt d}= \mathbf{0}  \\
 %      \bar{R}_{\tt e}\nabla_{x_{\tt e}}\mathcal{H}_{\tt d}= \mathbf{0} & \implies & \nabla_{x_{\tt e}}\mathcal{H}_{\tt d} = \mathbf{0}
 %  \end{array} \right.
%\end{equation*}
%Hence,
%\begin{equation*}
 %   \nabla_{x_{\tt e}}\mathcal{H}_{\tt d} = \mathbf{0} \iff \nabla\varphi(x_{\tt e}) + \nabla_{x_{\tt e}}\Phi_{2}(z) = \mathbf{0}.
%\end{equation*}
Thus, the result follows from LaSalle's invariance principle and \eqref{detect}.  $\square$ 
\end{pf*}
%\pbr{The mentioned approach captures the coupling between two mechanical and electrical subsystems through the desired interconnection matrix and Hamiltonian function. }
\begin{Rem}
The closed-loop systems obtained in Propositions \ref{pro:case1} and \ref{pro:case2} preserve the pH structure because \eqref{eq:pHcondition} is satisfied.
\end{Rem}
%You can mention that a (trivial) way to guarantee that (18) holds is setting D_{\tt d} as zero, i.e., excluding coupled damping in the controller. 

\begin{Rem}
 Assumption \ref{ass:class1} guarantees the existence of $\Phi_{1}(x_{\tt e})$ satisfying \eqref{phi11} and \eqref{phi12}. Similarly, Assumption \ref{ass:class2} ensures the existence of $\Phi_{2}(z)$ satisfying \eqref{Phic21} and \eqref{Phic22}. 
\end{Rem}

\section{Tracking control design}
\label{sec:track}

Inspired by Theorem \ref{th:contractive} and the timed IDA-PBC method in \cite{yaghmaei2017trajectory}, we design controllers that achieve exponential trajectory tracking for two specific classes of weakly coupled EM systems. Similar to the regulation design in Section \ref{sec:reg}, the proposed approaches circumvent the requirement of solving PDEs or implementing coordinate transformations. Moreover, the resulting controllers are static.
In particular, we propose target dynamics and energy function to avoid the need for solving PDEs or employing any change of coordinates. Therefore, based on the results reported in \cite{yaghmaei2017trajectory}, we suggest target dynamics that correspond to a contractive system to guarantee the convergence of trajectories in the closed-loop system. Such target dynamics include the pH structure but also more general contractive systems.
As mentioned before, we leverage the advantages of contraction theory to simplify the complexity of using Lyapunov analysis in proving exponential stability for both tracking and regulation problems for the EM systems characterized by \eqref{eq:open-loop elecmech}.

\subsection{Tracking for EM systems with only partial mechanical damping}
\label{sec:trackcase1}

The following assumption characterizes the class of EM systems studied in this section.

\begin{Asu}
\label{ass:class3}
The system \eqref{eq:open-loop elecmech} satisfies $R_{\tt m} \succ 0$ and 
\begin{equation}
	 \label{eq:cond1T2}
	\alpha I\prec \begin{bmatrix}
	\nabla_q^2 \mathcal{H}(\eta)  & \quad  \nabla_{q}(\nabla_{p}\mathcal{H}(\eta)) \\ \nabla_{p}(\nabla_{q} \mathcal{H}(\eta)) & \quad M^{-1}(q)  \end{bmatrix} \prec \beta I; \quad \forall \eta \in \mathcal{S},
\end{equation}
 where $\mathcal{S} \subseteq  R^{2n_{\tt m}+n_{\tt e}}$ is an open set containing the desired trajectory $\eta^\star$ and $0<\alpha<\beta$.
\end{Asu}

%In the following discussion, we specifically introduce the first class of weakly coupled EM systems.
%To align with the concept of the tracking design method, we define an open subset $\mathcal{S} \subseteq  R^{2n_{\tt m}+n_{\tt e}}$ containing the desired trajectory $\eta^\star(t)$.
%\label{sec:trackcase1}

From a mathematical perspective, Assumption \ref{ass:class3} is more restrictive than Assumption \ref{ass:class1}. However, a broad range of EM systems satisfying the latter also satisfy the former. 
Notably, for EM applications with a constant mass inertia matrix, like MEMS, this assumption reduces to Assumption \ref{ass:class1}, satisfying for all $ \eta \in \mathcal{S}$. 
% \pbr{In particular, for EM systems with a constant inertia matrix, Assumptions \ref{ass:class1} and \ref{ass:class3} are equivalent @@I think this is correct, but let's double-check@@.}

%This assumption includes the EM systems with a positive definite mechanical damping $R_{\tt m}$ and a stable convex energy function $H(\eta)$ for all $\eta \in \mathcal{S}$. Besides,
%in comparison to Assumption \ref{ass:class1}, the condition \eqref{eq:cond1T2} is needed to hold on the set $\mathcal{S}$. Notably, for applications with a constant mass inertia matrix, like MEMS, this assumption reduces to Assumption \ref{ass:class1}, satisfying for all $ \eta \in \mathcal{S}$.
The next theorem proposes a tracking method for EM systems satisfying Assumption \ref{ass:class3}.

\begin{Thm}
\label{the:track1}
Consider the system \eqref{eq:open-loop elecmech} satisfying Assumption \ref{ass:class3}, a twice differentiable function $\MAP{\Theta_1}{ \rea^{n_{\tt e}} \times \rea_{+} }{}$, and matrices $ K_{\tt{e}}  \succeq 0$, $D_{\tt d}$  such that:
\begin{itemize}
    \item Given $0<\alpha_{1}<\beta_{1}$, $\Theta_1 (x_{\tt e}, t)$ satisfies
	 \begin{equation}
	 \label{eq:cond2T2}
	\hspace{-3 mm} \alpha_{ 1} I\prec \nabla_{\eta}^2 \mathcal{H}(\eta)+\nabla_{\eta}^2 \Theta_1 (x_{\tt e}, t) \prec \beta_{ 1} I, \; \forall\eta \in  \mathcal{S}. 
	 \end{equation}
   %where $\alpha_{\tt 1},\beta_{\tt 1}>0$.
    \item  Set $\Gamma=\mathbf{0}$ in \eqref{Eq:R-J}, there exists $\varepsilon>0$ such that the matrix
	\begin{equation}
	\label{eq:cond3T2}
	\mathcal{N}_{1} \triangleq \begin{bmatrix}
	\mathcal{F}_{\tt d} & \left(  1-\frac{\alpha_{\tt 1}}{\beta_{\tt 1}} \right) \mathcal{F}_{\tt d}\mathcal{F}^\top_{\tt d}\\ -( 1-\frac{\alpha_{\tt 1}}{\beta_{\tt 1}}+\varepsilon)I & -\mathcal{F}^\top_{\tt d}
	\end{bmatrix},
	\end{equation}
has no eigenvalues on the imaginary axis.
    \item The next equality holds 
    \begin{equation}
	\label{eq:feasibleT2}
\begin{array}{rcl}
\dot{x}_{\tt{e}}^\star&&= D_{\tt d}M^{-1}(q^\star)p^\star\\&&+(J_{\tt{e}}-{\bar R}_{\tt{e}})\lbrace\Psi(q^\star)x^\star_{\tt{e}}+ \nabla_{x_{\tt{e}}}\Theta_1 (x^\star_{\tt{e}}, t)\rbrace.
\end{array}
\end{equation}

\end{itemize}
The static feedback controller
\begin{equation}
\label{eq:u1-track}
\begin{array}{rcl}
u &=& G_{\tt e}^{-1}\left\lbrace  D_{\tt d}M^{-1}(q)p-K_{\tt{e}}\Psi(q)x_{\tt{e}} \right.
\\&&+\left.\left( J_{\tt{e}}-{\bar R}_{\tt{e}}\right)\nabla_{x_{\tt{e}}}\Theta_1 (x_{\tt{e}}, t)\right\rbrace,
\end{array}
\end{equation}
ensure that the trajectories of the closed-loop system converge exponentially to $\eta^\star$.
\end{Thm}
\begin{pf*}{Proof.}
The proof consists in proving that the conditions in Theorem \ref{th:contractive} are satisfied by the closed-loop system. To this end, note that \eqref{eq:open-loop elecmech} in closed-loop with the controller \eqref{eq:u1-track} takes the form
\eqref{eq:closedloop} with 
  \begin{equation}
  \label{eq:HdClass3}
\Gamma=\mathbf{0},\quad\mathcal{H}_{\tt d}(\eta,t)=\mathcal{H}(\eta)+\Theta_1(x_{\tt e},t).
  \end{equation}
To assess the condition (i) of Theorem \ref{th:contractive}, we show there exist matrices $K_{\tt e}$ and $D_{\tt d}$ such that $\mathcal{F}_{\tt d}$ is Hurwitz. To this end, note that
\begin{equation*}
 \mathcal{F}_{\tt d} = \begin{bmatrix}
     \mathbf{0} & I & \mathbf{0} \\ -I & -R_{\tt m} & \mathbf{0} \\ \mathbf{0} & D_{\tt d} & J_{\tt e}-\bar{R}_{\tt e}
 \end{bmatrix},  
\end{equation*}
is a triangular block matrix. Consequently, $\mathcal{F}_{\tt d}$ satisfies condition (i) in Theorem \ref{th:contractive} if and only if the matrices $J_{\tt e}-\bar{R}_{\tt e}$ and
%for the corresponding closed-loop system \eqref{eq:closedloop}, we discuss the eigenvalues of $\mathcal{F}_{\tt d}$.
%  Accordingly, $\mathcal{F}_{\tt d}$ is Hurwitz if and only if the following matrix
\begin{equation*}
\bar B\triangleq \begin{bmatrix}
\mathbf{0} & \quad  I\\[2 mm]
-I & \quad -R_{\tt m}
\end{bmatrix},
\end{equation*}
are Hurwitz. In this regard, Bendixson's theorem---see \cite{bernstein2009matrix}---establishes that the upper bound of the eigenvalues of any real matrix is given by the maximum eigenvalue of its symmetric part. Consequently, any matrix $K_{\tt e}$ satisfying \eqref{barRe} guarantees that $J_{\tt e}-\bar{R}_{\tt e}$ is Hurwitz. Furthermore, $\bar{B}$ is the first companion matrix of the polynomial $L(\lambda)\triangleq I\lambda^2+R_{\tt m}\lambda+I$. Hence, the values of $\lambda$ such that $\det\{L(\lambda)\}=0$ corresponds to the eigenvalues of $\bar{B}$. Additionally, $R_{\tt m}$ is positive definite, implying that its singular value decomposition takes the form
\begin{equation}
    R_{\tt m} = U_{\tt m}^{\top}\Sigma_{\tt m}U_{\tt m}, \label{SVD}
\end{equation}
where $U_{\tt m}$ is an orthogonal matrix and $\Sigma_{\tt m} \in \rea^{n_{\tt m} \times n_{\tt m}}$ is diagonal and positive definite. Using \eqref{SVD}, we can define
  \begin{equation}
  \label{eq:L-landa}
   \tilde{L}(\lambda)\triangleq U_{\tt m}^\top L(\lambda)U_{\tt m}=I\lambda^2+\Sigma_{\tt m}\lambda+I.
   \end{equation}
Because $U_{m}$ is full rank, we have that
\begin{equation*}
\det\{L(\lambda)\}=0 \iff \det\{\tilde{L}(\lambda)\}=0.
\end{equation*}

Note that $\Sigma_{\tt m}$ is diagonal. Thus, it follows from \eqref{eq:L-landa} that the values of $\lambda$ that make the determinant of $L(\lambda)$ zero correspond to the roots of $n_{\tt m}$ second-order polynomials. However, all the coefficients of such polynomials are positive, implying that their roots are negative real numbers. Accordingly, $\bar{B}$ is Hurwitz.
%\begin{equation}
%\label{eq:detL}
%\begin{split}
%&\det\{L(\lambda)\}=\det\{\tilde{L}(\lambda)\}=\det(\lambda^{2} I+\Sigma_m\lambda+I)\\&=\prod_{\tt i=1}^{n_{\tt m}}(\lambda^2+\sigma_i\lambda+1).
%\end{split}
%\end{equation}

Note that the Hessian of $\mathcal{H}_{\tt d}(\eta,t)$ with respect to $\eta$ is given by
\begin{equation*}
 \begin{bmatrix}
	\nabla_q^2 \mathcal{H}  & \quad  \nabla_{q}(\nabla_{p}\mathcal{H}) & \quad \nabla_{q}(\nabla_{x_{\tt e}}\mathcal{H}) \\ \nabla_{p}(\nabla_{q} \mathcal{H}) & \quad M^{-1} & \mathbf{0} \\ \nabla_{x_{\tt e}}(\nabla_{q} \mathcal{H}) & \quad \mathbf{0} & \quad \nabla_{x_{\tt e}}^2 \mathcal{H}+\nabla_{x_{\tt e}}^2 \Theta_1   \end{bmatrix}.
\end{equation*}

Hence, Assumption \ref{ass:class3} guarantees the existence of $\Theta_1(\eta,t)$---for instance, a quadratic function---such that \eqref{eq:cond2T2} holds. Furthermore, given  \eqref{eq:cond2T2} and \eqref{eq:HdClass3}, implies that condition (ii) from Theorem \ref{th:contractive} is satisfied. Similarly, \eqref{eq:cond3T2} ensures that (iii) in Theorem \ref{th:contractive} holds.
%  Besides, considering Assumption \ref{ass:class3}, \eqref{eq:cond2T2} and \eqref{eq:cond3T2}  are satisfied \pbr{@why? If this is true, we need to modify the statement@}. Therefore, conditions (i), (ii), and (iii) in Theorem \ref{th:contractive} hold.
 Accordingly, the closed-loop system is contractive. Moreover, \eqref{eq:feasibleT2} $\eta^\star$ is a trajectory of the closed-loop system. Thereby, the convergence property in contractive systems \cite[Theorem 1]{lohmiller1998contraction} ensures that all the trajectories of the closed-loop system exponentially converge to $\eta^\star$. $\square$
\end{pf*}
\begin{Rem}
 Note that, in the proof of Theorem \ref{the:track1}, $D_{\tt d}$ does not need to satisfy \eqref{eq:pHcondition} to ensure that $\mathcal{F}_{\tt d}$ is Hurwitz. Thus, the coupled damping term does not need to preserve the pH structure to guarantee that the closed-loop system is contractive. In \cite{9804759}, the authors highlight that preserving the pH structure may lead to conservative results while injecting coupled damping. However, no general stability proof is provided when the pH structure is not preserved in the mentioned reference.
\end{Rem}

\subsection{Tracking for EM systems with only partial or no mechanical damping}
\label{sec:trackcase2}

As is the regulation case, we provide a controller for a class of EM systems with undamped or only partially damped mechanical subsystems below. We introduce the following assumption to characterize such systems.
% The second category of weakly coupled EM systems is presented in this subsection.  Following the tracking approach, we establish an open subset $\mathcal{Q} \subseteq  R^{2n_{\tt m}+n_{\tt e}}$ that encompasses the desired trajectory $\eta^\star(t)$.

\begin{Asu}
\label{ass:class4}
Given \eqref{eq:open-loop elecmech}, there exist an open set $\mathcal{Q} \subseteq  R^{2n_{\tt m}+n_{\tt e}}$ encompassing the desired trajectory $\eta^\star$ and a twice differentiable function $\MAP{\varphi_2}{\rea^{n_{\tt e}}}{}$ such that

\begin{equation}
\begin{array}{rcl}
 \nabla_{q}H_{\tt e}(q,x_{\tt e}) &=& -\Gamma\nabla\varphi_2(x_{\tt e}), \\
 \nabla^{2}\varphi_2(x_{\tt e}^{\star})&\succ& 0,
 \quad\forall \eta \in \mathcal{Q}.
\end{array} \label{assHe}
\end{equation}

% where $\Lambda_2 \in\rea^{n_{\tt m}\times n_{\tt e}}$ is a constant matrix.
\end{Asu}

Assumptions \ref{ass:class2} and \ref{ass:class4} are similar. However, the latter imposes more restrictions on the desired trajectory than the former, as the Hessian must be positive at every point of $\eta^{\star}$. The following theorem provides a control strategy to solve the trajectory-tracking problem for EM systems characterized by Assumption \ref{ass:class4}.

% Similar arguments in \textit{Assumption \ref{ass:class2}} apply to this case.
% The only difference is in satisfying
% the condition \eqref{assHe} within the subset $\mathcal{Q}$ rather than an equilibrium point.

% Now, considering the concept of contractive pH systems and the developed method in \textit{Proposition \ref{pro:case2}}, we suggest a tracking approach for the systems in \textit{Assumption \ref{ass:class3}}.

\begin{Thm}
\label{the:track2}
Consider \eqref{eq:open-loop elecmech} satisfying Assumption \ref{ass:class4}. Suppose there exist
% Suppose \eqref{eq:open-loop elecmech} under \textit{Assumption \ref{ass:class3}} and $\Gamma_2=\Lambda_2$.
a twice differentiable function $\MAP{\Theta_2}{\rea^{n_{\tt e}} \times \rea_{+}}{}$, $ K_{\tt{e}} \succ 0$ and $D_{\tt d} \in \rea^{n_{\tt e} \times n_{\tt m}}$ such that:
\begin{itemize}
 \item Conditions (i), (ii), and (iii) from Theorem \ref{th:contractive} hold for $\eta\in\mathcal{Q}$; $\mathcal{F}_{\tt d}=\mathcal{J}_{\tt d}-\mathcal{R}_{\tt d}$, with $\mathcal{J}_{\tt d}$ and $\mathcal{R}_{\tt d}$ defined in \eqref{Eq:R-J};
  \begin{equation}
  \label{eq:HdClass4}
      \mathcal{H}_{\tt d}(\eta,t)=H_1(q,p)+\varphi_2(x_{\tt e})+\Theta_2(z,t),
  \end{equation}
 with $H_1(q,p)\triangleq\frac{1}{2}{p^{\top}} M^{-1}(q)p+V(q)$  and $z$ is defined as in \eqref{z}.
 \item Given the desired trajectory $\eta^{\star}$ and $z^\star$ defined in \eqref{zstar},
    \begin{equation}
	\label{eq:L2}
 \begin{array}{rcl}
\dot{x}_{\tt{e}}^\star&=&(D_{\tt d}-\Gamma^\top) M^{-1}(q^\star) p^\star +(J_{\tt e} - \bar{R}_{\tt e})\nabla\varphi_2(x^\star_{\tt e})\\&-&\nabla_{z}\Theta_2 (z^\star, t).\vspace{-5 mm}\end{array}
\end{equation}
 \end{itemize}
% \begin{itemize}
%     \item The matrix $\mathcal{F}_{\tt d}$ in \eqref{Eq:R-J} is Hurwitz.
%     \item The following inequality holds:
% 	 \begin{equation}
% 	 \label{eq:cond22}
% 	 \alpha_{\tt 2} I\prec \nabla_\eta^2 H_1(q,p)+\nabla_\eta^2 \Theta_2 (z, t)+
%   \nabla_\eta^2 \varphi_2(x_{\tt e})
%   \prec \beta_{\tt 2} I,\quad \eta \in \mathcal{Q}
% 	 \end{equation}
%    where $H_1(q,p)\triangleq\frac{1}{2}{p^{\top}} M(q) ^{-1}p+V(q)$, $\alpha_{\tt 2},\beta_{\tt 2}>0$ and $z$ is presented in \eqref{z}.
%     \item The matrix
% 	\begin{equation}
% 	\label{eq:cond32}
% 	\mathcal{N}_{\tt 2} = \begin{bmatrix}
% 	\mathcal{F}_{\tt d} & \left(  1-\frac{\alpha_{\tt 2}}{\beta_{\tt 2}} \right) \mathcal{F}_{\tt d}\mathcal{F}_{\tt d}^\top\\ -( 1-\frac{\alpha_{\tt 2}}{\beta_{\tt 2}}+\varepsilon)I & -\mathcal{F}_{\tt d}^\top
% 	\end{bmatrix},
% 	\end{equation}
%     has no eigenvalues on the imaginary axis for a positive constant $\varepsilon$.
%     \item The next condition holds:
%     \begin{equation}
% 	\label{eq:L2}
%  \begin{array}{rcl}
% \dot{x}_{\tt{e}}^\star=-\Lambda^\top_2 M(q^\star) ^{-1}p^\star +(J_{\tt e} - \bar{R}_{\tt e})\nabla\varphi(x^\star_{\tt e})\\-\nabla_{z}\Theta_2 (z^\star, t).
%  \end{array}
% \end{equation}
% where $z^\star$ is denoted in \eqref{zstar}.
% \end{itemize}
% Hence, the static feedback controller
\vspace{3 mm}
The control law
 \begin{equation}
 \label{tracker2}
 \begin{array}{rcl}
  u &=& G_{\tt e}^{-1}\left\lbrace  \left( R_{\tt e}-J_{\tt e} \right)\Psi(q)x_{\tt{e}}+(D_{\tt d}- \Gamma^\top) M^{-1}(q)p\right. \\ &&+\left.\left(  J_{\tt e}-\bar{R}_{\tt e}\right)  \nabla\varphi_2(x_{\tt e})   \right\rbrace-\nabla_{z} \Theta_2(z,t), \vspace{-5 mm}
 \end{array}
 \vspace{3 mm}
\end{equation}

guarantees that, on $\mathcal{Q}$, the trajectories of the closed-loop system converge exponentially to $\eta^{\star}$.
% makes the dynamics \eqref{eq:open-loop elecmech}  a local exponential tracking controller for $\eta^\star$.\\
\end{Thm}
\begin{pf*}{Proof.}
 Given \eqref{z}, we have that
\begin{equation}
   \begin{array}{rcl}
      \nabla_{q}\Theta_2  & = & - \Gamma(J_{\tt e}-\bar{R}_{\tt e})^{-1}G_{\tt e}\nabla_{z}\Theta_2,  \\
      \nabla_{x_{\tt e}}\Theta_2  & = & -(J_{\tt e}-\bar{R}_{\tt e})^{-1}G_{\tt e}\nabla_{z}\Theta_2.
   \end{array}
\end{equation}

Accordingly,
\begin{equation}
    \begin{array}{rcl}
        \nabla_{z}\Theta_2 & = & -G_{\tt e}^{-1}(J_{\tt e}-\bar{R}_{\tt e})\nabla_{x_{\tt e}}\Theta_2\\
         \nabla_{q}\Theta_2  &=& \Gamma\nabla_{x_{\tt e}}\Theta_2.
    \end{array}\label{nTheta22}
\end{equation}

Given Assumption \ref{ass:class4} and \eqref{nTheta22}, the matching equation \eqref{eq:matching} is satisfied. Note that Assumption \ref{ass:class4} guarantees the existence of
$\Theta_2(z,t)$ such that
conditions (ii) from Theorem \ref{th:contractive} is satisfied.
Thus, \eqref{eq:open-loop elecmech} in closed loop with \eqref{tracker2} yields \eqref{eq:closedloop}. Moreover, because the conditions from Theorem \ref{th:contractive} are satisfied, the closed-loop system is contractive. Furthermore, \eqref{eq:L2} guarantees that $\eta^\star$ is a trajectory of the closed-loop system. Accordingly, all trajectories of the closed-loop system exponentially converge to $\eta^\star$ due to the convergence property in contractive systems. $\square$
\end{pf*}
% is derived with the following parameters:
%   \begin{equation}
%   \label{eq:HdClass4}
%       \Gamma=\Lambda_2,\,\,\ \mathcal{H}_{\tt d}(\eta,t)=H_1(q,p)+\varphi_2(x_{\tt e})+\Theta_2(z,t),
%   \end{equation}
%  Considering \textit{Assumption \ref{ass:class4}} and \eqref{nTheta22}, the matching equation \eqref{eq:matching} holds.
%   Since $\mathcal{F}_{\tt d}$ is Hurwitz , \eqref{eq:cond22} and \eqref{eq:cond32}  are satisfied, the conditions in \textit{Theorem \ref{th:contractive}} hold. Hence, the closed-loop system is contractive.
%   Besides, \eqref{eq:L2} guarantees $\eta^\star(t)$ to be one of the trajectories of the closed-loop system. Accordingly, all trajectories of the closed-loop system exponentially converge to $\eta^\star(t)$ due to the convergence property in contractive systems.
%\njm{Remark about z}

We stress that
the results of Theorems \ref{the:track1} and \ref{the:track2} require $\mathcal{F}_{\tt d}$ to be Hurwitz. However, preserving the pH structure is not necessary.  Additionally, the convergence properties of contractive systems are exponential. Therefore, by considering $\eta^{\star}$ as a constant configuration, the controllers proposed in Theorems \ref{the:track1} and \ref{the:track2} ensure exponential stability of the desired equilibrium.
Regarding the regulation problem, these controllers are less conservative compared to the controllers developed in Section \ref{sec:reg} because they do not need to satisfy the condition \eqref{eq:pHcondition}. Therefore, the coupled damping term can be set within a wider range without being restricted by pH structure resulting from \eqref{eq:pHcondition}.

\begin{Rem}
The stability results of Propositions \ref{pro:case1} and \ref{pro:case2} and Theorems \ref{the:track1} and \ref{the:track2} are global if the corresponding closed-loop energy function $\mathcal{H}_{\tt d}(\eta)$ is radially unbounded.
\end{Rem}

\section{Coupled damping and performance assessment}
\label{sec:coupleddamp}

The mechanical subsystem governs the transient performance of EM systems of the form \eqref{eq:open-loop elecmech}. Accordingly, it is crucial to provide adequate mechanical damping injection to shape the transient performance and improve the convergence rate of the closed-loop system. However, the control input only affects the dynamics of the electrical subsystem, making it impossible to modify the mechanical damping directly. To circumvent this problem, coupled damping can ameliorate the coupling between mechanical and electrical subsystems. In particular, coupled damping allows injecting damping into the mechanical subsystem from the electrical subsystem, improving the convergence rate and transient performance of the closed-loop system.

Note that, from \eqref{Eq:R-J}, 
 $\mathcal{J}_{\tt d}$ and $\mathcal{R}_{\tt d}$ couple the dynamics of the electrical states with the dynamics of $p$ through the coupling terms $\Gamma$ and $D_{\tt d}$. This coupling is not necessary to guarantee stability. However, it affects the closed-loop performance, as indicated in the following remark and the remainder of this subsection.

% From , we deduce that $\mathcal{J}_{\tt d}$ and $\mathcal{R}_{\tt d}$ couple the electrical subsystem, \emph{i.e.}  the electrical state $x_{\tt e}$, with the mechanical subsystem, \emph{i.e.} the momenta $p$

%  Note that we consider $\Gamma=0$ in \textit{Proposition \ref{pro:case1}} and \textit{Theorem \ref{the:track1}}. Therefore, $D_{\tt d}$ is the only term to increase the coupling between the subsystems.
%  As a result, in the case of weakly coupled EM systems with non-zero and low natural mechanical damping $R_{\tt m}$, where the system encounters insufficient mechanical damping, $D_{\tt d}$ plays an effective role in the transient behavior.
\begin{Rem}
Setting $D_{\tt d}=\mathbf{0}$ in the stabilization Propositions \ref{pro:case1} and \ref{pro:case2}, and in tracking Theorems \ref{the:track1} and \ref{the:track2}, we ensure the asymptotic and exponential stability of the closed-loop system, respectively. However, adopting a non-zero $D_{\tt d}$ significantly influences the shaping of the transit behavior in the closed-loop systems.
\end{Rem}
% \subsection{Convergence rate in contractive systems}
% \label{sec:CEM moddel}
% In this part, we investigate the convergence rate of the closed-loop system captured from the contraction property. Accordingly,  we develop a detailed understanding of the coupled damping effect on the system response.

To illustrate the effect of coupled damping on the convergence rate of the closed-loop system, we introduce the following lemma, whose proof can be found in \cite {lohmiller1998contraction}.\\
% In the following lemma, we characterize the contraction analysis, introduced in \cite {lohmiller1998contraction}, for the closed-loop system \eqref{eq:closedJ-R}.
\begin{lem}
\label{lem:contraction}
%and $\mathcal{H}_{\tt d}$ denoted in \eqref{eq:HdClass3} and \eqref{eq:HdClass4}, respectively.
Given the closed-loop system \eqref{eq:closedJ-R}. Suppose there exists a constant and nonsingular square matrix $\omega\in\rea^{2n_{\tt m}+n_{\tt e}\times 2n_{\tt m}+n_{\tt e}}$ such that
the following equation holds on a region of the state space denoted as $\mathcal{B}_{\tt c}$:
  \begin{equation}
  \label{eq:crate}
  \hspace{-3 mm}   \omega \mathcal{F}_{\tt d}\nabla^2_{\tt \eta}\mathcal{H}_{\tt d}(\eta,t)\omega^{-1}
     +\omega^{-\top}  \nabla_{\tt\eta}^2 \mathcal{H}_{\tt d}(\eta,t) \mathcal{F}_{\tt d}^\top \omega^{\top} \preceq -	\sigma I,
  \end{equation}
  where the constant $\sigma$ is positive. Then, the region $\mathcal{B}_{\tt c}$ is called a contraction region with respect to the metric $\Omega:=\omega^\top \omega$ for the system \eqref{eq:closedJ-R}. Therefore, any trajectory of the system \eqref{eq:closedJ-R}, which starts in a ball of constant radius with respect to the metric $\Omega$, centered at a given trajectory and contained at all times in $\mathcal{B}_{\tt c}$, remains in that ball and converges exponentially to this trajectory.
  \end{lem}

%\begin{proof}  
%Considering the closed-loop system (16), the proof immediately follows from  Definition 2 and Theorem 2 presented in \cite[Section 3]{lohmiller1998contraction} \pbr{@It would be nice if we could point out the formal proof. However, I couldn't find it in \cite {lohmiller1998contraction}. \njm{the proof has presented through the paper. I guess that the mentioned theorems are known enough in Contraction area.}}
%\end{proof}

In the above lemma, $\sigma$ denotes the convergence rate of the closed-loop system \eqref{eq:closedJ-R}. The following remark provides intuition into how coupled damping terms affect the convergence rate.

\begin{Rem}
 Define the matrix
 \begin{equation*}
  \Xi \triangleq \omega \mathcal{F}_{\tt d}\nabla^2_{\tt \eta}\mathcal{H}_{\tt d}(\eta,t)\omega^{-1}.
 \end{equation*}
 Then, \eqref{eq:crate} can be rewritten as
 \begin{equation}
  \Xi + \Xi^{\top}\preceq -\sigma I,
 \end{equation}
 which implies from Bendixson's theorem that the real part of the eigenvalues of $\Xi$ is negative. Moreover, a suitable $\sigma$ is given by the largest eigenvalue of the symmetric part of $\Xi$, which must be negative definite. However, such an eigenvalue is modified by the presence of coupled damping. To note that define
 \begin{equation*}
 \mathcal{D}_{\tt d}\triangleq \begin{bmatrix}
                                \mathbf{0} & \mathbf{0} & \mathbf{0} \\ \mathbf{0} & \mathbf{0} & \mathbf{0} \\ \mathbf{0} & D_{\tt d} & \mathbf{0}
                               \end{bmatrix},
\end{equation*}
and let $\Xi_{1}$ correspond to the case $D_{\tt d}=\mathbf{0}$ in $\mathcal{F}_{\tt d}$ and $\Xi_{2}$ to $D_{\tt d}\neq\mathbf{0}$ in $\mathcal{F}_{\tt d}$. Hence, we obtain the relation
\begin{equation*}
 \Xi_{2} = \Xi_{1}+\omega \mathcal{D}_{\tt d}\nabla^2_{\tt \eta}\mathcal{H}_{\tt d}(\eta,t)\omega^{-1},
\end{equation*}
where
\begin{equation*}
 \mbox{eig}\{\Xi_{2}\}= \mbox{eig}\{\Xi_{1}+\omega \mathcal{D}_{\tt d}\nabla^2_{\tt \eta}\mathcal{H}_{\tt d}(\eta,t)\omega^{-1}\}\neq
 \mbox{eig}\{\Xi_{1}\}.
\end{equation*}

Therefore, the largest eigenvalue of the symmetric part of $\Xi_2$ is affected by the matrix $\mathcal{D}_{\tt d}$, and thereby coupled damping $D_{\tt d}$.
 \end{Rem}
 
To analyze the effect of coupled damping on the convergence rate more in detail, below, we adopt the rationale used in the proof of Theorem \ref{th:contractive} provided in \cite{yaghmaei2017trajectory} to characterize the contraction property exposed in Lemma \ref{lem:contraction}.

Given \eqref{eq:closedJ-R}, we assume the existence of $0<\alpha_{3}<\beta_{3}$ and $\mathcal{B}_{\tt c}  \subseteq R^{2n_{\tt m}+n_{\tt e}} $ such that
  \begin{equation}
	 \label{eq:assbound}
	 \alpha_{ 3} I\prec \nabla_{\eta}^2 \mathcal{H}_{\tt d}(\eta,t) \prec \beta_{ 3} I, \quad \eta \in \mathcal{B}_{\tt c}.
	 \end{equation}
Using \eqref{eq:assbound} and the property
 \begin{equation*}
     A\prec B \implies CAC^\top\prec CBC^\top,
 \end{equation*}
 where $A, B,$ and $C$ are square matrices, we obtain 
   \begin{eqnarray}
   \label{eq:manipulation}
       (\omega \mathcal{F}_{\tt d}+\omega^{-\top})\nabla_{\tt\eta}^2 \mathcal{H}_{\tt d}(\eta,t)
       ( \mathcal{F}^\top_{\tt d}\omega^\top+\omega^{-1})
       \preceq \nonumber \\
       \beta_3 (\omega \mathcal{F}_{\tt d}+\omega^{-\top})
       ( \mathcal{F}^\top_{\tt d} \omega^\top+\omega^{-1}),
   \end{eqnarray}
because $\nabla_{\tt\eta}^2 \mathcal{H}_{\tt d}(\eta,t)\succ 0$, by considering the left side of the inequality \eqref{eq:assbound} and performing some algebraic manipulations in \eqref{eq:manipulation} and noting that  $-\omega^{-\top}\nabla_{\tt\eta}^2 \mathcal{H}_{\tt d}(\eta,t)\omega^{-1}\preceq -\omega^{-\top} \alpha_3 \omega $, we conclude that 
  \begin{eqnarray}
  \label{eq:bound}
       \omega \mathcal{F}_{\tt d} \nabla_{\tt\eta}^2 \mathcal{H}_{\tt d}(\eta,t) \omega^{-1}+ \omega^{-\top}  \nabla_{\tt\eta}^2 \mathcal{H}_{\tt d}(\eta,t) \mathcal{F}_{\tt d}^\top \omega^{\top} \preceq \nonumber  \\
     \beta_3 \omega^{-\top}\big(\Omega \mathcal{F}_{\tt d}+
      \mathcal{F}^\top_{\tt d} \Omega +\gamma I+\gamma  \Omega \mathcal{F}_{\tt d} \mathcal{F}^\top_{\tt d}  \Omega \big)
      \omega^{-1},
  \end{eqnarray}
  where $\gamma\triangleq1-\frac{\alpha_3}{\beta_3}$.  Accordingly, the right side of \eqref{eq:bound} is negative definite if the following Riccati equation has as a positive definite solution $\Omega$ for some $\varepsilon>0$:
    \begin{eqnarray}
  \label{eq:riccati}
\Omega \mathcal{F}_{\tt d}+
      \mathcal{F}^\top_{\tt d} \Omega +(\gamma+\varepsilon) I+\gamma  \Omega \mathcal{F}_{\tt d} \mathcal{F}^\top_{\tt d}  \Omega=0.
  \end{eqnarray}
  
Hence, replacing \eqref{eq:riccati} in \eqref{eq:bound} yields
   \begin{eqnarray}
 \label{eq:bound2}
       \omega \mathcal{F}_{\tt d} \nabla_{\tt\eta}^2 \mathcal{H}_{\tt d}(\eta,t) \omega^{-1}+ \omega^{-\top}  \nabla_{\tt\eta}^2 \mathcal{H}_{\tt d}(\eta,t) \mathcal{F}_{\tt d}^\top \omega^{\top} \preceq \nonumber  \\
 -\beta_{3} \varepsilon \omega^{-\top}
 \omega^{-1},
 \end{eqnarray}
 
Because $\omega$ is invertible, there exists an upper bound such that $-\beta_3 \varepsilon \omega^{-\top}\omega^{-1}  \preceq  -\beta_3 \varepsilon\lambda_{max}\{ \omega^{-\top}\omega^{-1}\} I $. From \eqref{eq:crate} and \eqref{eq:bound2}, we deduce that the system \eqref{eq:closedJ-R} is contractive and $\sigma\triangleq\beta_3 \varepsilon\lambda_{max}\{ \omega^{-\top}\omega^{-1}\}$ is the convergence rate of the system. Notably, $\sigma$ depends on the value of $\varepsilon$, which is derived from the Riccati equation \eqref{eq:riccati}.
Therefore, considering two cases, $D_{\tt d} = 0$ and $D_{\tt d} \neq 0$, in $\mathcal{F}_{\tt d}$ leads to different values of $\varepsilon$ in \eqref{eq:riccati}. Hence, the coupled damping $D_{\tt d}$ changes the rate of convergence in \eqref{eq:closedJ-R}.

\section{Simulation}
\label{sec:ex}
In this section, we apply the controllers proposed in sections \ref{sec:reg} and \ref{sec:track} to two different EM systems and study the closed-loop performance with and without coupled damping.
\subsection{Micro electro-mechanical optical switch}

Here, we explore the design of the controllers to address regulation and tracking problems in an optical switch system. The system dynamics are reported in \cite{borja2016constructive} and \cite{borovic2004control}. This model can be represented by the pH system \eqref{eq:open-loop elecmech} with $n_{\tt e}=n_{\tt m}=1$, $J_{\tt e}=0$, $R_{\tt e}=\frac{1}{r_{\tt e}}$,  $G_{\tt e}=\frac{1}{r_{\tt e}}$, where $r_{\tt e}>0 $, the constant mass inertia $m$
 and the following parameters:   
\begin{equation*}
    V=\frac{1}{2}a_1q^2+\frac{1}{4}a_2q^4,\,\ H_{\tt e}=\frac{x_{\tt e}^2}{2c_1(q+c_0)},\,\
    \Psi=\frac{1}{c_1(q+c_0)},
\end{equation*}
where the spring constants are $a_1,a_2>0$, and the capacitor constants are $c_0, c_1>0$.
In the subsequent subsections, we provide regulation and trajectory-tracking control approaches. For both approaches,  the following numerical values are employed as the system parameters:
\\
$c_0=15\times10^{-6},\quad
c_1=35.6\times10^{-9}, \quad
m=2.35\times10^{-9},\\
a_1=0.46,\quad
a_2=0.0973,\quad
R_{\tt m}=5.5\times10^{-7},\quad
r_{\tt e}=100$.
\subsubsection{Regulation}
The control aim is to stabilize the system at the desired position $q_{\tt d}$. The desired  equilibrium \eqref{eq:desired eq} for this system is
\begin{eqnarray}
\label{eq:}
\eta_{\tt d}=\big(q_{\tt d}, 0, 
 (c_0+q_{\tt d})\sqrt{2c_1q_{\tt d}(a_1+a_2q^2_{\tt d})}\big),
\end{eqnarray}
where $q_{\tt d}(a_1+a_2q^2_{\tt d})>0$.
Notably,
\begin{eqnarray}
\label{eq:gradH}
  \nabla^{2}_{q}\mathcal{H}(q_{\tt d})=a_1+3a_2q_d^2+d^2_1d_2,  
\end{eqnarray}
where $d_1=\sqrt{2c_1q_{\tt d}(a_1+a_2q^2_{\tt d})}$ and $d_2=\frac{1}{c_1(c_0+q_{\tt d})}$ , is positive for the system's parameters. Therefore, the Assumption \ref{ass:class3} is satisfied for the system.
Accordingly, we use Theorem \ref{the:track1} to design a regulation controller for the system with the following closed-loop parameters:
\begin{eqnarray}
\label{eq:closedPrameterreg}
K_{\tt e}=\frac{1}{\bar{r}_{\tt e}}-\frac{1}{{r}_{\tt e}}, \quad \theta_1 (x_{\tt e}) =\frac{1}{2}(x_{\tt e}-{x_{\tt e}}_{\tt d}+L_1)^2,
%\Phi_1(x_{\tt e})=\frac{1}{2}(x_{\tt e}-\kappa)^2
\end{eqnarray}
where $\bar r_{\tt e}>0$, and from \eqref{eq:feasibleT2} we have
\begin{eqnarray}
\label{eq:RegL}
L_1=-\frac{{x_{\tt e}}_{\tt d}}{c_1(q_{\tt d}+c_0)}.
\end{eqnarray}
%and $\kappa={x_{\tt e}}_{\tt d}+\frac{{x_{\tt e}}_{\tt d}}{c_1(q+c_0)}$. 
%Besides, the condition \eqref{phi12} is evaluated as follows:
%\begin{eqnarray}
%\emph{A}=\begin{bmatrix}
%	a_1+3a_2q^2_{\tt d}+d^2_1d_2&  -d_1d_2\\[1.8 mm] -d_1d_2 & d_2+1
%	\end{bmatrix},
%\end{eqnarray}

Besides, the Hessian of $\mathcal{H}_{\tt d}$ in \eqref{eq:HdClass3} with respect to $\eta$ is given by
\begin{equation}
\label{eq:Hessian}
 \begin{bmatrix}
	a_1+3a_2q^2_{\tt d}+d^2_1d_2  & \mathbf{0} & -d_1d_2 \\ \mathbf{0} & \quad M^{-1} & \mathbf{0} \\ -d_1d_2 & \quad \mathbf{0} & \quad d_2+1  \end{bmatrix}.
\end{equation}

Therefore, Assumption \ref{ass:class3} guarantees  that \eqref{eq:cond2T2} holds.
Moreover, the condition \eqref{eq:cond3T2} holds through the appropriate selection of the closed-loop parameters.
%\njm{By choosing the appropriate values for parameters $\bar r_{\tt e}$ and $D_{\tt d }$, we ensure that the condition \eqref{phic122} is satisfied.}

Thereby, we conclude from \eqref{eq:u1-track} and \eqref{eq:RegL} as  the next controller
\begin{equation}
\label{eq:C2reg}
 u = -\frac{r_{\tt e}}{\bar r_{\tt e}} ( x_{\tt e}-{x_{\tt e}}_{\tt d}+L_1 )- K_{\tt e}r_{\tt e}\frac{x_{\tt{e}}}{{c_1(q+c_0)}}+r_{\tt e}\frac{D_{\tt d}}{m}p,
\end{equation}
stabilizes the system at the equilibrium point $\eta_{\tt d}$. 
In \eqref{eq:closedPrameterreg}, $\Theta_1(x_{\tt e})$ does not depend explicitly on $t$. Consequently, $\mathcal{H}_{\tt d}(\eta)$ in \eqref{eq:HdClass3} does not (explicitly) depend on $t$. This is because we apply the control approach outlined in Theorem \ref{the:track1} to address the regulation problem for a constant desired reference. 

For the simulation, we choose the following numerical values:
\begin{eqnarray}
 \label{eq:numrug}
 q_{\tt d}= 3 \times 10^{-5},
 \quad\bar{r}_{\tt e}=100.
\end{eqnarray}

\begin{figure}
\centering
\includegraphics[width=1\columnwidth]{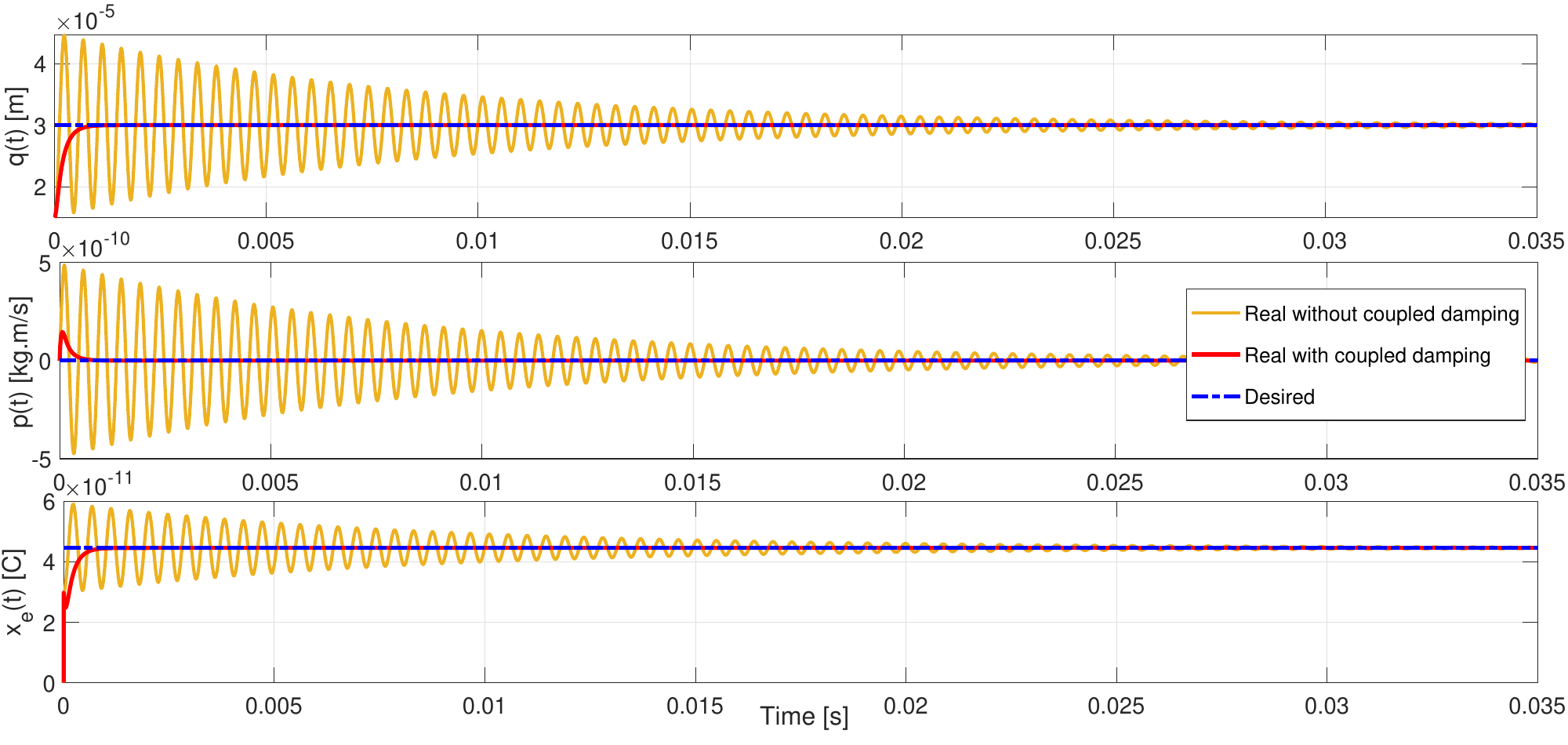}
\centering
\vspace{-0.3cm}
\caption{$q(t)$ converges to the desired position $q_{\tt d}$ for the  initial condition $[15\times 10^{-6},\ 0,\ 0 ]^\top$ via  control \eqref{eq:c1} without and with the coupled damping, respectively.}
\label{fig:regulation}
\end{figure}

 To study the impact of the coupled damping, we simulate the results for both scenarios:
without coupled damping (setting $D_{\tt d}=0$) and
with coupled damping (setting $D_{\tt {d}}=-1$). In Fig. \ref{fig:regulation}, the orange lines depict the first scenario with $D_{\tt d}=0$, where
the mechanical and electrical states
of the system converging to the equilibrium point as time approaches infinity.
However, it is evident that the transient responses include oscillations before reaching
to $\eta_{\tt d}$.
To avoid these oscillations and modify the transient
responses, we leverage the functionality of the coupled damping concept, as elaborated in Section \ref{sec:coupleddamp}. Hence, fixing $D_{\tt d}=-1$, Fig. \ref{fig:regulation} indicates that the system's state trajectory, shown in red, converges to $\eta_{\tt d}$, exhibiting no oscillations. 
%\pbr{@@If you want to show the use of Proposition 1, then remove the coupled damping from here and move it to the next subsection, where you can comment on the extra benefits of Theorem 2 (i.e., more flexibility to inject coupled damping). If you decide to do that, I would rename the following section as trajectory-tracking and regulation with coupled damping injection.}
Note that Theorem \ref{the:track1} and Proposition \ref{pro:case1} can be used for solving the regulation problem in this system. However, here, we apply the controller proposed in Theorem \ref{the:track1}---hence, we set $q_{\tt d} = q_{\star}$---because it is less conservative concerning the coupled damping that can be injected into the system. In particular, it is not needed to satisfy \eqref{eq:pHcondition}, i.e., preserving the pH structure is not required.
%\njm{Note that we use the controller proposed in Theorem \ref{the:track1}, which does not need to satisfy \eqref{eq:pHcondition}.  It makes it less conservative with respect to various numerical options compared to the controller in Proposition \ref{pro:case1}.}
\subsubsection{Trajectory Tracking}
The control objective here is for the position $q$ to track the desired trajectory $q^\star$, yielding
\begin{eqnarray}
&p^\star=m\dot{q}^\star,\\&
x^\star_{\tt e}=(c_0+q^\star)\sqrt{2c_1(m\Ddot{q}^\star+b\dot{q}^\star+a_1q^\star+a_2{q^\star}^3)},
\end{eqnarray}
where $(m\Ddot{q}^\star+b\dot{q}^\star+a_1q^\star+a_2{q^\star}^3)>0$.
The Assumption \ref{ass:class3} is fulfilled by the system dynamics. Hence, we use {Theorem \ref{the:track1}} to design a tracker for the system. Accordingly, we select $K_{\tt e}$ as in \eqref{eq:closedPrameterreg} and the following parameter:
\begin{eqnarray}
\label{eq:closedPrameter}
\theta_2 (x_{\tt e}, t) =\frac{1}{2}(x_{\tt e}-x^\star_{\tt e}+L_2(t))^2,
\end{eqnarray}
where $\MAP{L_2}{\rea_{+}}{}$.
Subsequently, from \eqref{eq:feasibleT2}, we obtain
\begin{eqnarray}
L_2(t)=-\bar{r}_{\tt e}\dot{x}^\star_{\tt e}-\frac{x^\star_{\tt e}}{c_1(q^\star+c_0)}+\bar r_{\tt e}\frac{D_{\tt d}}{m}p^\star.
\end{eqnarray}

Thereby, the controller is derived from \eqref{eq:u1-track} and \eqref{eq:closedPrameter} as 
\begin{eqnarray}
\label{C1}
u=-\frac{r_{\tt e}}{\bar r_{\tt e}}(x_{\tt e}-x^\star_{\tt e}+L_2(t))
-\frac{K_{\tt e}r_{\tt e}x_{\tt e}}{c_1(q+c_0)}
+r_{\tt e}\frac{D_{\tt d}}{m}p.
 \end{eqnarray}

 According to \eqref{eq:Hessian}, \eqref{eq:cond2T2} and \eqref{eq:cond3T2} hold through the appropriate selection of the closed-loop parameters. The numerical values for tracking aim are chosen as
 \begin{eqnarray}
 \label{eq:csim}
 &q^\star=0.05+0.05\sin{(30 t)},\quad\label{eq:reference}
 \bar{r}_{\tt e}=100.
  \end{eqnarray}

\begin{figure}
\centering
\includegraphics[width=1\columnwidth]{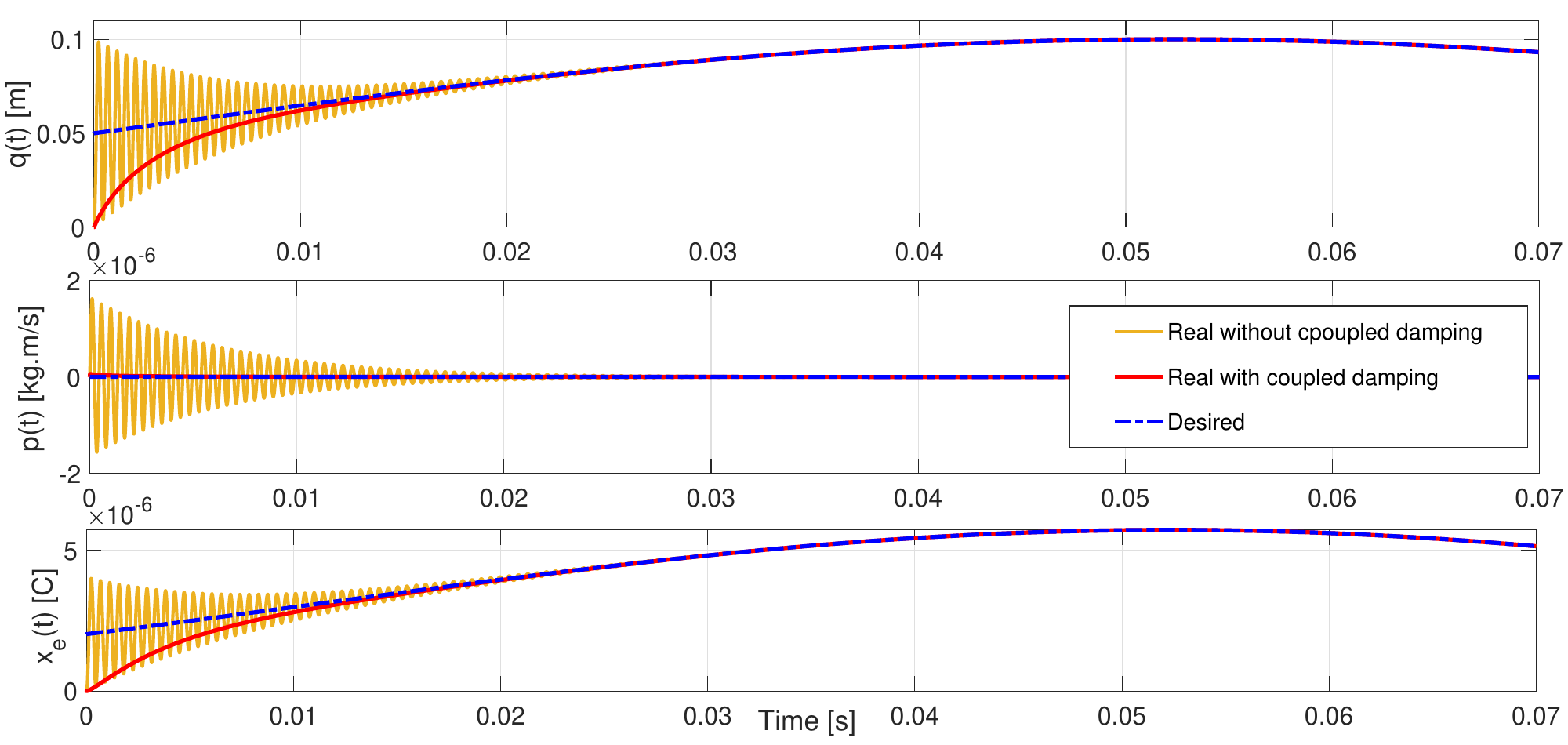}
\centering
\vspace{-0.3cm}
\caption{$q(t)$ exponentially tracks the signal $q^\star(t)$ for the  initial condition $[15\times 10^{-6},\ 0,\ 0 ]^\top$ via tracking control \eqref{eq:u1-track} without and with the coupled damping, respectively . }
\label{fig:tracking}
\end{figure}

Similar to the regulation part, we consider two scenarios: without coupled damping (setting $D_{\tt d}$ = 0) and with coupled damping (setting
$D_{d}=-0.4$).
 The real trajectory without coupled damping (the orange lines) in Fig. \ref{fig:tracking} shows the mechanical and electrical states of the system converge to the desired reference over time with oscillating responses.
 To remove these oscillations and modify the transient responses, we refer to the coupled damping concept in Section \ref{sec:coupleddamp}. Accordingly, by considering $D_{\tt d}=-0.4$, the mechanical and electrical states trajectories with coupled damping (the red lines in Fig. \ref{fig:tracking})  track exponentially $q^\star$ in \eqref{eq:reference} in Fig. \ref{fig:tracking}.
Besides, it indicates that the control method in \eqref{C1} effectively shapes transient performance without oscillations while inserting damping into the mechanical subsystem through coupled damping. %\pbr{@@depending on space and your opinion, you can add a plot of $q$ following the desired trajectory for a longer time (maybe 0.3 seconds). This is optional}
\subsection{Magnetic levitation System}

In this part, we show the effectiveness of {Theorem \ref{the:track2}} in addressing the trajectory tracking problem for magnetic levitation (or maglev) systems \cite{rodriguez2000novel}.
The dynamics of the magnetic levitation system is represented as the pH system \eqref{eq:open-loop elecmech} with $G_{\tt e}=1$, $J_{\tt e}=0$, the constant mass inertia $m$ and the dimension of the states $n_{\tt e}=n_{\tt m}=1$. The electrical energy \eqref{eq:elec energy} and the mechanical potential energy are defined as follows
\begin{equation}
\label{open-loop mech1}
\arraycolsep=1pt
\def\arraystretch{1.2}
\begin{array}{rcl}
&
 H_{\tt e}(q,x_{\tt e})=\frac{1}{2k}(c-q)x^2_{\tt e},\,\,\, V(q)=bq,            \end{array}
\end{equation}
where $b,c,k>0$. It is required that $q<c$. The control aim is to track the desired trajectory $q^\star$, yielding
\begin{eqnarray*}
&p^\star=m\dot{q}^\star, \quad
x^\star_{\tt e}=\sqrt{2k(m\Ddot{q}^\star+b)},
\end{eqnarray*}
where $(m\Ddot{q}^\star+b)>0$
. The {Assumption \ref{ass:class4}} is satisfied for the system dynamics with $\Gamma=\frac{1}{2k}$ and $\nabla \varphi_2=x_{\tt e}^2$. Therefore, according to {Theorem \ref{the:track2}}, we choose $\Theta_2(z,t)$ as follows
\begin{equation}
 K_{\tt e}=\bar{R}_{\tt e}-R_{\tt e}\quad,\Theta_2(z,t)=\frac{k_c}{2}(z-L_3(t))^2,
\end{equation}
where $\bar{R}_{\tt e}$, $k_{\tt c}>0$ and $\MAP{L_3}{\rea_{+}}{}$.
The condition \eqref{eq:L2} holds with the following function
\begin{equation}
L_3(t)=\frac{1}{k_{\tt c}}\lbrace \dot x^\star_{\tt e}+\bar R_{\tt e} {x^\star_{\tt e}}^2+k_{\tt c}z^\star+\frac{1+D_{\tt d}}{2km}p^\star\rbrace,
\end{equation}
where $z^\star$ in \eqref{zstar}. Accordingly, the controller  given in \eqref{tracker2} takes the form
\begin{equation}
\label{eq:controlmaglv}
    u=\frac{-1+D_{\tt d}}{2km}p -\bar{R}_{\tt e} x^2_{\tt e}
    -k_c(z-L_3(t))
    +\frac{R_{\tt e}}{k}(c-q)x_{\tt e},
\end{equation}

Hence, by appropriately selecting values of the parameters $\bar K_{\tt e}$ and $k_{\tt c}$ %\pbr{@@remember that you select $k_c$, not $\bar R_{\tt e}$}, 
conditions (i)--(iii) in Theorem \ref{th:contractive} are satisfied.
For simulation, we use the following numerical values:
\begin{equation*}
\begin{array}{cl}
   & k=6.4042 \times 10^{-1} \mbox{N m/A},\quad R_{\tt e}=2.25,\quad c=0.005 \mbox{m},
   \\& b=0.8280  \mbox{kg m/s},
   \quad m=0.0844 \mbox{kg}, \quad k_{\tt c}=20\\&
    \end{array}
\end{equation*}

\begin{figure}
\centering
\includegraphics[width=1\columnwidth]{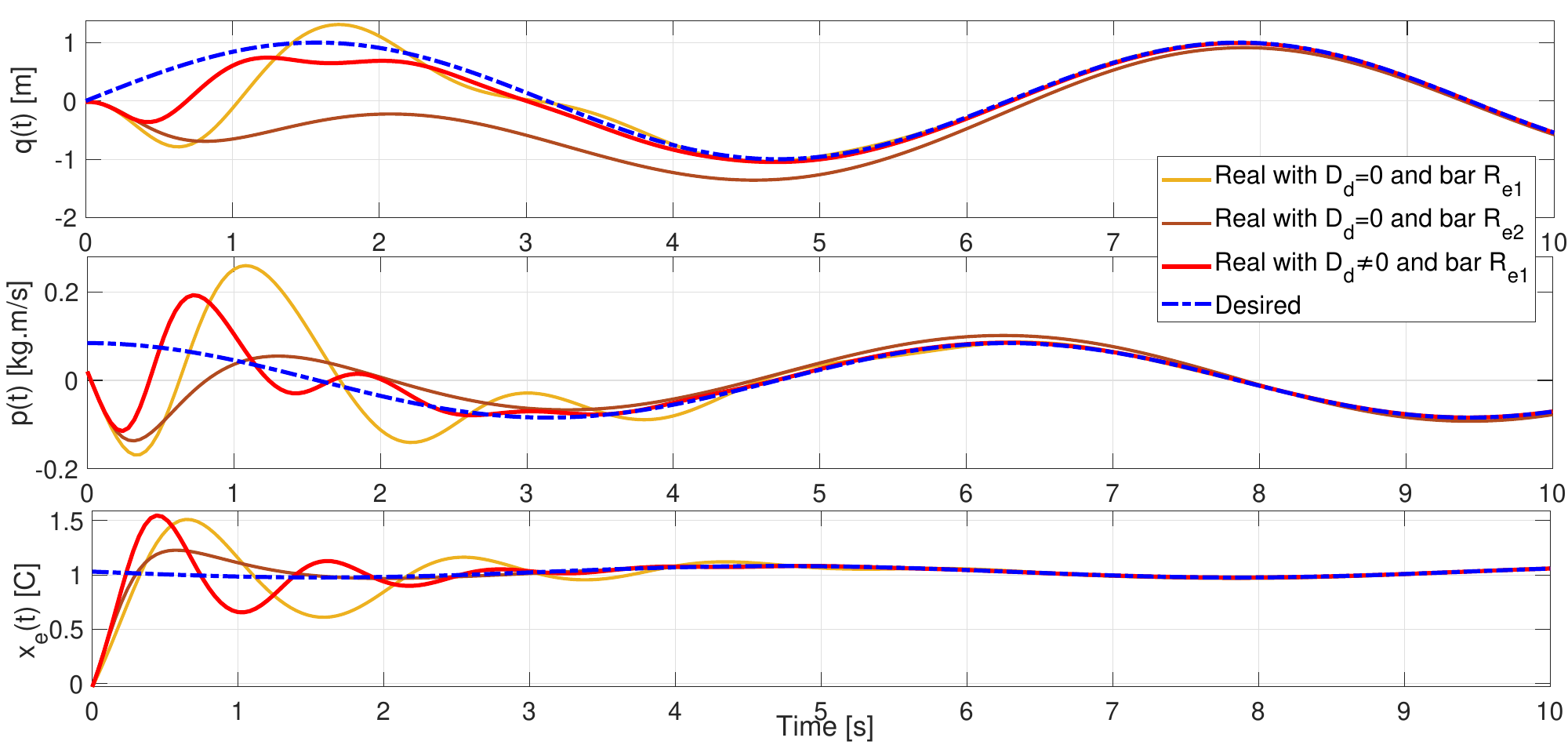}
\centering
\vspace{-0.3cm}
\caption{$q(t)$ exponentially tracks the signal $q^\star(t)$ via tracking controller \eqref{eq:controlmaglv} without and with the coupled damping, and different electrical damping ($\bar R_{\tt e1}<\bar R_{\tt e2}$).}
\label{fiq:maglvtraj}
\end{figure}

To examine the impact of coupled damping, various scenarios are simulated with different values of $D_{\tt d}$ and $\bar R_{\tt e}$ in Fig. \ref{fiq:maglvtraj}. We set the parameters: $D_{\tt d}=-1, \,\  \bar R_{\tt e1}=0.82, \,\  \bar R_{\tt e2}=2.82$. As shown in Fig. \ref{fiq:maglvtraj}, all trajectories converge to the desired one as a result of employing the proposed controller \eqref{eq:controlmaglv}, with different transient behaviors. Notably, the position trajectory $q(t)$ exhibits fewer oscillations when coupled damping is injected (the red lines) in comparison to scenarios without coupled damping (the orange lines). Note that by increasing the electrical damping ($\bar R_{\tt e1}<\bar R_{\tt e2}$), the trajectories tend towards an overdamped behavior, where the convergence rate is remarkably slow, particularly in the $q$ trajectory. Therefore, inserting the damping into the mechanical subsystem through the coupled damping is more effective than increasing the electrical one. 

\begin{figure}
\centering
\includegraphics[width=1\columnwidth]{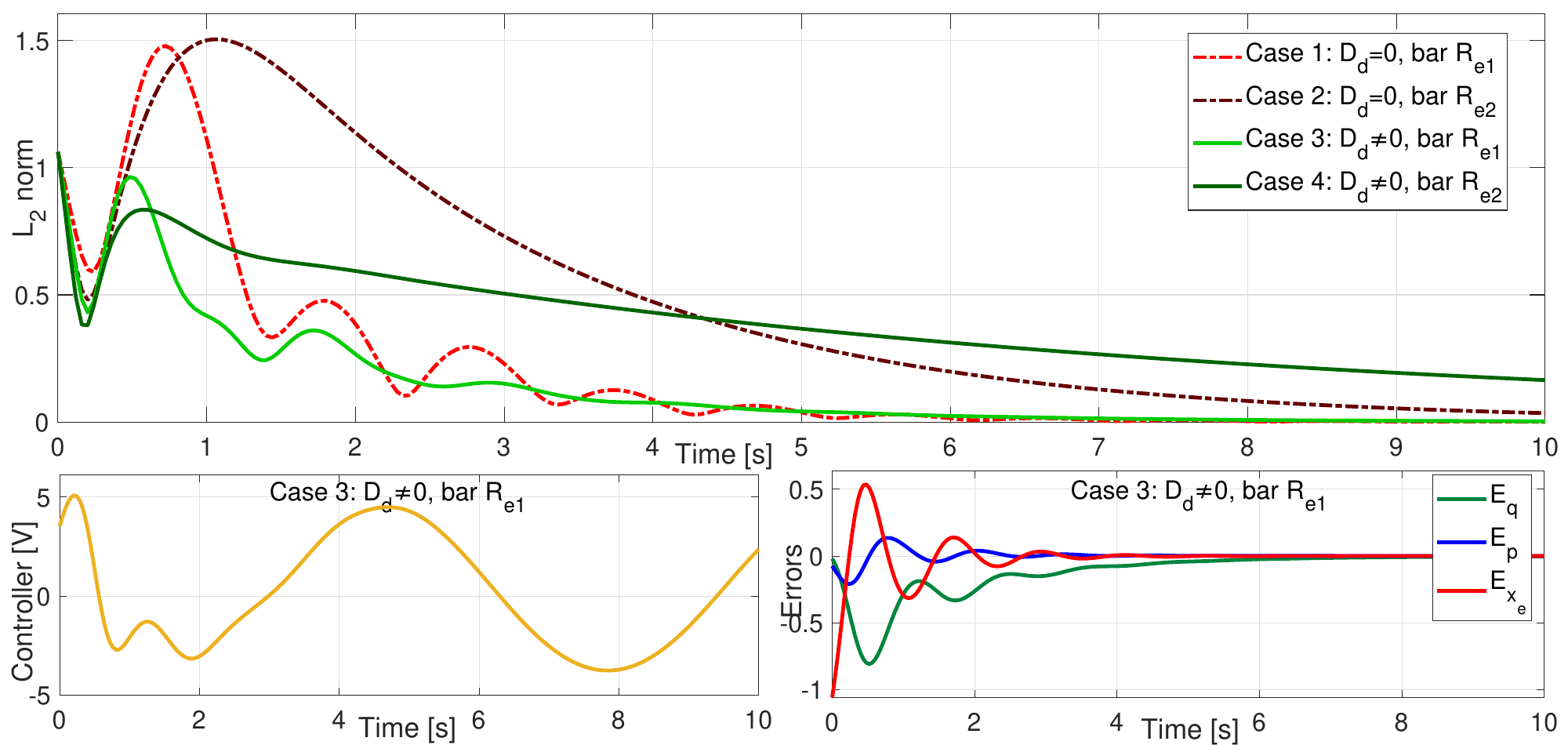}
\centering
\vspace{-0.3cm}
\caption{$\mathcal{L}_2$-norms of the error for four different cases (upper plot). The control signal plot (bottom left) corresponds to Case 3. The error signals are depicted in the plot at the bottom right.}
\label{maglavnorm}
\end{figure}

For a detailed comparison, Fig. \ref{maglavnorm} shows the $\mathcal{L}_2$-norm of the error for the simulation of different cases. In this figure, we observe that the error and the corresponding $\mathcal{L}_2$-norm approach zero in all the cases. However, in Cases 2 and 4, corresponding to larger electrical damping $\bar R_{\tt e2}$, the convergence is slower than in Cases 1 and 3. Interestingly, Cases 3 and 4---which involve coupled damping---demonstrate a reduction in the magnitude of the norm during the initial time intervals. Hence, we deduce that coupled damping improves the performance of the closed-loop system in terms of the $\mathcal{L}_2$-norm of the error. We stress that, in Case 3,  where coupled damping is injected and the electrical damping is chosen as $\bar R_{{\tt e}1}$, there is a significant improvement in the error norm in terms of oscillations and magnitude. The plot of the control signal in Fig. \ref{maglavnorm} corresponds to this case.

%\pbr{@and?? What's the purpose of these plots? What do you observe? What is relevant to the reader? Why only one control signal and not the 4 you considered?}

\section{Concluding Remarks}
\label{sec:con}

In this paper, we addressed the regulation and trajectory tracking problems for two classes of weakly coupled EM systems. We formulated these systems within a unified pH framework suitable for a broad class of weakly coupled systems. These systems are characterized by assumptions that are not restrictive and can be easily verified. Subsequently, we developed control methods based on Lyapunov and contraction theories. Moreover, we explored the use of coupled damping injection to enhance the transient performance and convergence rate of the closed-loop system. Our results provide evidence of reduced oscillations in the transient performance of applications, such as micro-electro-mechanical optical switches, through the application of coupled damping.

\bibliographystyle{plain}        % Include this if you use bibtex 
\bibliography{autosam} 

          % bib file to produce the bibliography
                             % with bibtex (preferred)
                                                   
%\begin{thebibliography}{xx}  % you can also add the bibliography by hand

%\bibitem[Able(1956)]{Abl:56}
%B.C. Able.
%\newblock Nucleic acid content of microscope.
%\newblock \emph{Nature}, 135:\penalty0 7--9, 1956.

%\bibitem[Able et~al.(1954)Able, Tagg, and Rush]{AbTaRu:54}
%B.C. Able, R.A. Tagg, and M.~Rush.
%\newblock Enzyme-catalyzed cellular transanimations.
%\newblock In A.F. Round, editor, \emph{Advances in Enzymology}, volume~2, pages
%  125--247. Academic Press, New York, 3rd edition, 1954.

%\bibitem[Keohane(1958)]{Keo:58}
%R.~Keohane.
%\newblock \emph{Power and Interdependence: World Politics in Transitions}.
%\newblock Little, Brown \& Co., Boston, 1958.

%\bibitem[Powers(1985)]{Pow:85}
%T.~Powers.
%\newblock Is there a way out?
%\newblock \emph{Harpers}, pages 35--47, June 1985.

%\bibitem[Soukhanov(1992)]{Heritage:92}
%A.~H. Soukhanov, editor.
%\newblock \emph{{The American Heritage. Dictionary of the American Language}}.
%\newblock Houghton Mifflin Company, 1992.

%\end{thebibliography}

\end{document}